\documentclass[prb,showpacs,twocolumn,amsmath,amssymb,floatfix]{revtex4-1}
\usepackage{graphicx}
\usepackage{bm}
\usepackage{bbm}
\usepackage{color}
\usepackage{amsmath}
\usepackage{amssymb}
\usepackage{color}
\usepackage{footnote}
\usepackage{epstopdf}
\usepackage{hyperref}
\usepackage{appendix}
\usepackage[applemac]{inputenc}

\begin{document}
\title{Odd-frequency superconducting pairing and subgap density of states at the edge of a two-dimensional topological insulator without magnetism}
\author{Jorge Cayao and Annica M. Black-Schaffer}
\affiliation{Department of Physics and Astronomy, Uppsala University, Box 516, S-751 20 Uppsala, Sweden}
\date{\today} 
\begin{abstract}
We investigate the emergence and consequences of odd-frequency spin-triplet $s$-wave pairing in superconducting hybrid junctions at the edge of a two-dimensional topological insulator without any magnetism.
More specifically, we consider several different normal-superconductor hybrid systems at the topological insulator edge, where spin-singlet $s$-wave superconducting pairing is proximity-induced from an external conventional superconductor.
We perform fully analytical calculations and show that odd-frequency mixed spin-triplet $s$-wave pairing arises due to the unique spin-momentum locking in the topological insulator edge state and the naturally non-constant pairing potential profile in hybrid systems. 
Importantly, we establish a one-to-one correspondence between the local density of states (LDOS) at low energies and the odd-frequency spin-triplet pairing in NS, NSN and SNS junctions along the topological insulator edge; at interfaces the enhancement in the LDOS can directly be attributed to the contribution of odd-frequency pairing. Furthermore, in SNS junctions we show that the emergence of the zero-energy LDOS peak at the superconducting phase $\phi=\pi$ is associated purely with odd-frequency pairing in the middle of the junction.

\end{abstract}
\maketitle
\section{Introduction}
\label{sect0}
Superconductivity is strongly characterized by the Cooper pair wave function, or the pairing amplitude, obeying the antisymmetry condition imposed by Fermi-Dirac statistics, which restricts its spin and spatial symmetries. The antisymmetry condition also allows for the emergence of odd-frequency superconductivity, where the pairing amplitude is odd in the time, or equivalently frequency, parameter.
The existence of odd-frequency superconductivity was first postulated by Berezinskii\cite{bere74} in 1974, when he pointed out that there is no symmetry restriction for the existence of odd-frequency spin triplet $s$-wave pairing in $^{3}$He. 
Subsequent works first took this idea to analyze spin-triplet $s$-wave superconductivity in disordered systems\cite{PhysRevLett.66.1533,PhysRevB.46.8393,PhysRevB.60.3485} and later extended this suggestion to proposed odd-frequency superconductivity in bulk systems with spin-singlet $p$-wave pairing.\cite{PhysRevB.45.13125,PhysRevB.52.1271,0953-8984-9-2-002} More recently, bulk odd-frequency superconductivity has also been found in multiband superconductors\cite{PhysRevB.88.104514,PhysRevB.92.094517,PhysRevB.92.224508} and in superconductors subjected to a time-dependent drive.\cite{PhysRevB.94.094518, triola17}

Following another route for non-bulk materials, it has been shown that odd-frequency superconductivity also emerges as an induced effect in hybrid systems where superconductivity is induced by proximity effect. For ferromagnet-superconductor (FS) junctions it is now well established, both theoretically and experimentally, that it is odd-frequency spin triplet $s$-wave pairing that explains the long-range proximity effect into the ferromagnet even when using conventional spin-singlet $s$-wave superconductors.\cite{PhysRevLett.86.4096,PhysRevLett.90.117006,RevModPhys.77.1321,PhysRevB.73.104412,PhysRevB.75.104509, 7870d3ff91ed485fa3e55e901ff81c80, 0953-8984-26-45-453201, 0034-4885-78-10-104501, EschrigNat15} In these systems the spin-rotation symmetry is broken by the ferromagnet, which allows for the formation of the spin-triplet state. 
Afterwards, it was also shown that normal metal-superconductor (NS) junctions also exhibit odd-frequency pairing. 
In this case, however, a conventional superconductor can only induce odd-frequency spin-singlet $p$-wave pairing, 
since only translational symmetry is broken.\cite{PhysRevLett.98.037003,PhysRevLett.99.037005,Eschrig2007} 
However, according to Anderson's theorem\cite{Anderson-theorem} only $s$-wave pairing is intrinsically stable against 
ever present non-magnetic disorder, and therefore this latter odd-frequency $p$-wave is much less stable than the 
odd-frequency $s$-wave state present in FS junctions. Moreover, normal metals are also fragile to disorder due to 
Anderson localization, namely they allow for finite elastic backscattering processes from non-magnetic impurities, 
causing dissipation of electric current. 
Odd-frequency spin-triplet $s$-wave pairs can however appear at interfaces of diffusive normal metals if the external superconductor is unconventional with spin-triplet $p$-wave symmetry.\cite{PhysRevLett.98.037003} 
More recently, the appearance of odd-frequency pairing was also found in systems with 
Rashba spin-orbit coupling.\cite{PhysRevB.92.134512,PhysRevB.95.184518}

On the other hand, topological insulators,\cite{RevModPhys.82.3045,RevModPhys.83.1057,Ando13} characterized by having an insulating bulk but metallic surface states, have strongly suppressed backscattering in their surface states. In fact, for two-dimensional topological insulators (2DTI) backscattering is completely absent due to the perfect helicity of the edge states. Induced $s$-wave superconductivity in a 2DTI edge therefore represents a promising and very disorder-insensitive platform for the search of robust odd-frequency pairing.
Indeed, it has recently been demonstrated that the combination of surface state helicity and a finite in-surface gradient in a proximity-induced spin-singlet $s$-wave superconducting state gives rise to odd-frequency spin-triplet $s$-wave pairing.\cite{PhysRevB.86.144506,PhysRevB.87.220506} It is the helicity of the TI surface state that allows a spatial symmetry breaking (finite in-surface gradient) to be effectively converted into a change of the spin symmetry, such that spin-triplet $s$-wave pairing is generated. Notably this mechanism does not need any presence of ferromagnetic regions or even a magnetic field, and thus does not destroy the topological protection of the TI surface states. 

There exits now a growing body of work focusing in detail on behavior of odd-frequency pairing in superconducting hybrid systems at both the surface of 3DTIs\cite{bo2016,PhysRevB.92.205424} and edge of 2DTIs.\cite{PhysRevB.92.100507} However, in these cases finite ferromagnetic regions has been assumed, which destroys the topological protection of the TI surface states. For the situation of odd-frequency pairing without magnetism in 2DTIs there exists no detailed analysis of the pairing amplitudes and especially not of their relationship to the local density of states (LDOS).
In this work we fill this gap and investigate in detail NS, NSN, and SNS hybrid junctions at the edge of a 2DTI without any magnetism present. Our study is based on retarded Green's functions extracted from scattering states and allow us to analytically extract all pairing amplitudes as well as the LDOS in the junctions. 
Very generally, we show that breaking translation symmetry at the NS interface(s) gives rise to four different symmetry classes at 
the NS interfaces: Even-frequency, spin-Singlet, Even in space (ESE), i.e.~the conventional order, Odd-frequency, spin-Singlet, Odd in space (OSO), Even-frequency spin-Triplet, Odd in space (ETO), and Odd-frequency, spin-Triplet, Even in space (OTE). In the absence of magnetism in the system we only generate mixed spin-triplet pairing, such that $m_z = 0$ for the Cooper pairs. Disorder stability is moreover preferential for the local $s$-wave states present in the ESE and OTE classes and we therefore focus primarily on these. These local pairing terms are also naturally the contributions most directly connected with the LDOS which is also a local quantity. 
A strong relationship between LDOS and  odd-frequency pairing has been previously already established in systems formed out of normal metals with unconventional superconductors\cite{PhysRevB.76.054522,PhysRevLett.98.037003,Nagaosa12} or when magnetism is present\cite{PhysRevB.75.134510,PhysRevB.87.104513,PhysRevB.92.014508,PhysRevB.92.205424}. With this work we extend this relationship also to 2DTI superconducting hybrid junctions without any magnetism.

In both NS and NSN junction we find strongly dominating OTE pairing at very low energies in the superconducting interface region, with an exponential decay into the bulk of S. In fact, at zero energy the ESE contribution is completely suppressed and only OTE pairing is present. We also find that the LDOS in the S region experience the same frequency dependence and exponential decay into the bulk of S as the OTE pairing. This shows that the very low energy contribution to the LDOS arises entirely from the OTE pairing. Moreover, we also find that the conductance in NS junctions follows the same behavior as the OTE component at low energies, suggesting that the main contribution to the conductance is also of OTE nature.
Our results here are strongly connected to the fact that the Andreev reflection magnitude at the NS interface reaches its maximum for energies within the gap, a unique characteristic of the helical edges of 2DTIs.\cite{PhysRevB.82.081303,PhysRevLett.109.186603}

In SNS junctions the LDOS reveals the formation of Andreev bound states (ABSs) for energies within the energy gap of 
the superconductor. We find that the ESE and OTE pairing magnitudes exactly capture their emergence, but they behave 
very differently for different values of the superconducting phase $\phi$ across the junction. Indeed, for very short 
junctions we obtain that at $\phi=0$ the ESE pairing dominates over a completely suppressed OTE contribution. 
At $\phi=\pi$ the ESE is instead zero while the OTE becomes dominant and even exhibits a zero-energy peak just as 
the LDOS thanks to the topologically protected zero-energy crossing of the ABSs. Moreover, the 
supercurrent across short junctions exhibits its maximum value at $\phi=\pi$, as a result of the 
resonant zero-energy peak which possesses purely OTE pairing.
In longer junctions, the increasing number of ABSs within the gap is also exactly captured in the ESE and OTE amplitudes. In fact, in the middle part of the N region the situation is very similar to the short junction case with only OTE pairing present at $\phi=\pi$. Here the ABSs again have protected crossings generating resonant peaks in the LDOS, which exactly correspond to resonant peaks in the OTE pairing. Thus the LDOS signatures of the ABSs at $\phi = \pi$ is entirely a consequence of OTE pairing. 

On the experimental side, HgCd/HgTe\cite{PhysRevLett.96.106802,Bernevig06,konig07,Roth09} and InAs/GaSb\cite{PhysRevLett.100.236601,PhysRevLett.107.136603} heterostructures represent two of the most promising 2DTI. In both cases, induced superconductivity has already been demonstrated.\cite{Yacoby14,vlad15}
Moreover, the superconducting junction geometries we consider comprise of a realistic platform for both LDOS and conductance measurements, as has also been demonstrated in experiments. \cite{PhysRevLett.109.186603,Yacoby14,vlad15,Bocquillon17} Therefore, all experimental prerequisites already exist for the systems we study.
 
This paper is organized as follows. In Sec.\,\ref{sect1} we present the model and the method based on retarded Green's functions calculated from scattering states. In Sec.\,\ref{sec2} we perform a detailed analysis of the pairing amplitudes and investigate their strong relationship with the LDOS for subgap energies for NS, NSN, as well as SNS junctions along the 2DTI edge. We present our conclusions in Sec.\,\ref{concl}. For completeness, in Appendices \ref{AppG},\,\ref{NSApp},\,\ref{NSNApp}, and\,\ref{SNSApp} we provide a detailed account of the method and analytical calculations reported in this work.

\section{Model and method}
\label{sect1}
\begin{figure*}[!ht]
\begin{minipage}[t]{\linewidth}
\centering
\includegraphics[width=.99\textwidth]{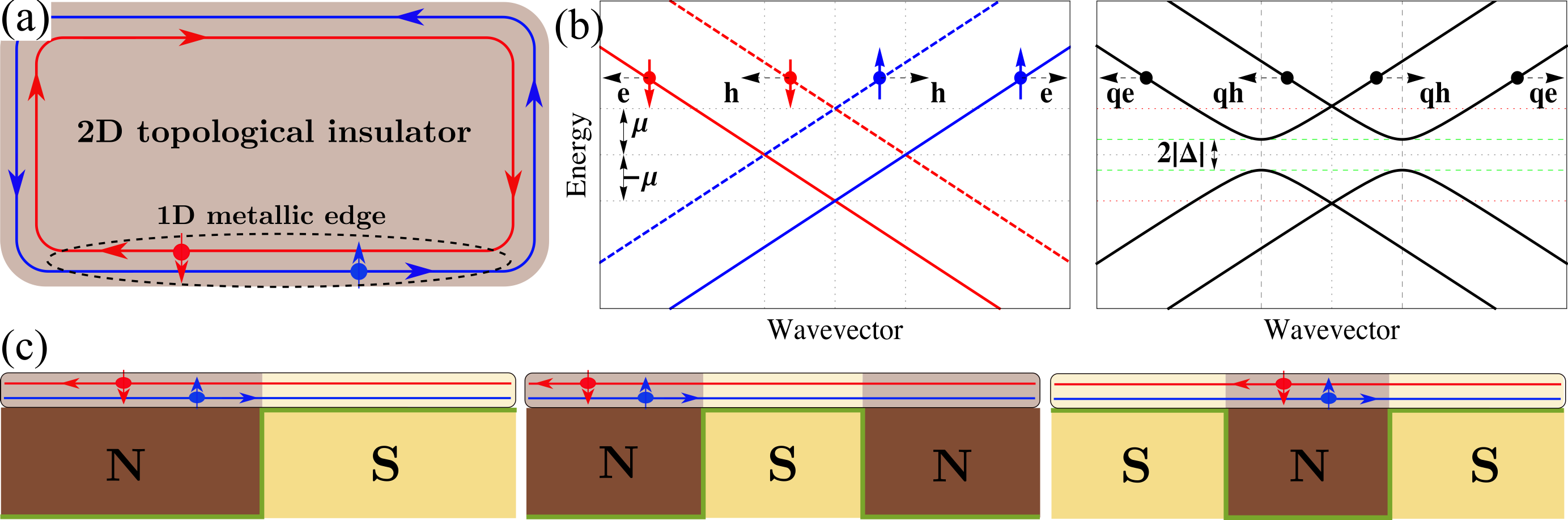} 
\caption{(Color online) (a) Schematic 2DTI, where the edge forms a 1D metallic system with counter-propagating edge modes carrying opposite spin (blue and red filled circles with arrows indicating spin direction). (b) In the normal state N (left) the metallic edge states exhibit a linear dispersion, while an $s$-wave superconductor S (right) induces a pairing potential $\Delta$ into the edge, which mixes the electron and hole branches by opening a gap at the Fermi momenta. Circles indicate the possible scattering processes at a SN interface for electron (e), hole (h), quasielectron (qe), and quasihole (qh) particles with horizontal arrows indicating propagation direction.
(c) Hybrid junction configurations: NS (left), NSN (middle), SNS (right) when the 1D metallic edge is partially proximitized to an external superconductor. Green lines show the profile of $\Delta$ through the junctions.}
\label{fig1}
\end{minipage}
\end{figure*}

We consider a 2D topological insulator (2DTI), with its 1D metallic edge in proximity to a conventional spin-singlet $s$-wave superconductor. This system is modeled by the Bogoliubov-de Gennes (BdG) Hamiltonian
\begin{equation}
\label{H2DSC}
\begin{split}
H_{BdG}&= v_{F}p_{x}\tau_{z}\sigma_{z}-\mu\tau_{z}\sigma_{0}+{\bf \Delta}(x)\tau_{x}\sigma_{0}\,,
\end{split}
\end{equation}
in the basis $ \Psi(x)=
 (\psi_{\uparrow},
 \psi_{\downarrow},
 \psi_{\downarrow}^{\dagger},
 -\psi_{\uparrow}^{\dagger}
 )^{T}$, where $T$ denotes the transpose operation, $\psi^{\dagger}_{\sigma}(x)$ adds an electron with spin $\sigma=\uparrow,\downarrow$ at position $x$ along the edge. 
 The first term represents the 1D metallic edge of a 2DTI,\cite{PhysRevLett.95.226801,PhysRevB.79.161408,0034-4885-75-7-076501} where the spin quantization direction is along the $z$-axis, $p_{x}=-i\partial_{x}$, $v_{F}$ is the velocity of the edge state, $\sigma_{i}$ is the $i$th Pauli spin matrix and $\tau_{i}$ is the $i$th Pauli matrix in Nambu space. The chemical potential is represented by $\mu$ and determines the filling. 
In the last term ${\bf \Delta}(x)=\Delta(x)\,{\rm e}^{i\phi}$ is the induced superconducting pairing potential at the edge. A good interface enables this pairing potential to be induced into the region(s) of the metallic edge in contact with an external superconductor as a result of proximity effect.\cite{DeGennes,Doh272} Notice that the pairing potential $\Delta(x)$ generally depends on the position at the edge in hybrid junctions. In the normal state (N) the edge states exhibit a linear energy versus momentum dispersion with $\Delta = 0$, while in superconducting (S) regions a finite $\Delta$ mixes the electron and hole branches and opens a gap at the Fermi momenta as shown in Fig.\,\ref{fig1}(a,b).
 
We will consider NS, NSN and SNS junctions formed at the edge of the 2DTI, such that the profile of the induced pairing potential is approximated by the step-like functions indicated in green in Fig.\,\ref{fig1}(c). The NS junction captures the physics of the N-S interface, while the NSN and SNS configurations represent finite S regions and Josephson junctions, respectively. 
Here the length, $L_{\rm N}$ and $L_{\rm S}$, respectively, of the middle regions become important parameters. Also, in the SNS junction it is possible to attain a finite phase difference between the two S regions, with pairing potentials $\Delta_{\rm L}=\Delta$ and $\Delta_{\rm R}=\Delta\,{\rm e}^{i\phi}$, respectively.

\subsection{Retarded Green's function}
\label{RGF}
In this work we primarily investigate the pairing amplitudes and LDOS in different hybrid junctions. 
For this we follow the formalism based on retarded and advanced Green's functions, corresponding to outgoing and incoming waves, respectively. 
In general, the retarded Green's function can be calculated using\cite{PhysRev.175.559,0034-4885-63-10-202} 
\begin{equation}
\label{GFUNCTION}
\begin{split}
&G^{r}(x,x',\omega)=\\
&\begin{cases}
   \alpha_{1} \Psi_{1}(x)\tilde{\Psi}_{3}^{T}(x')+\alpha_{2} \Psi_{1}(x)\tilde{\Psi}_{4}^{T}(x') &\\
   + \alpha_{3} \Psi_{2}(x)\tilde{\Psi}_{3}^{T}(x')+\alpha_{4} \Psi_{2}(x)\tilde{\Psi}_{4}^{T}(x'), & x>x', \\
   \beta_{1} \Psi_{3}(x)\tilde{\Psi}_{1}^{T}(x')+  \beta_{2} \Psi_{4}(x)\tilde{\Psi}_{1}^{T}(x') &\\
   +   \beta_{3} \Psi_{3}(x)\tilde{\Psi}_{2}^{T}(x')+  \beta_{4} \Psi_{4}(x)\tilde{\Psi}_{2}^{T}(x'), & x<x',
\end{cases}
\end{split}
\end{equation}
where $\Psi_{i}$, $i=1,2,3,4$ are the wave functions representing the four scattering processes at the NS interface at $x = x'$ indicated by black arrows in Fig.\,\ref{fig1}(b), i.e.~right-moving electron, right-moving hole, left-moving quasi-electron, and left-moving quasi-hole. $\tilde{\Psi}_{i}$ corresponds to the conjugated processes obtained using $\tilde{H}_{BdG}(k)=H_{BdG}^{*}(-k)=H_{BdG}^{T}(-k)$ instead of Eq.\,(\ref{H2DSC}). The detailed form of these wave functions is provided in the appendix for NS, NSN, and SNS junctions.
The retarded Green's function satisfies the equation of motion
\begin{equation}
[\omega-H_{BdG}(x)]G^{r}(x,x',\omega)=\delta(x-x')\,,
\end{equation}
where its integration around $x=x'$,
\begin{equation}
\label{discontiG}
\Big[G^{r}(x>x')\Big]_{x=x'}-\Big[G^{r}(x<x')\Big]_{x=x'}=\frac{\sigma_{z}\tau_{z}}{iv_{f}}\,,
\end{equation}
provides a system of equations that allows us to find the coefficients $\alpha_{i}$ and $\beta_{i}$ in Eq.\,(\ref{GFUNCTION}). Once the retarded Green's functions are found, the advanced Green's function can be calculated using $G^{a}(x,x',\omega)=[G^{r}(x',x,\omega)]^{\dagger}$. 

The Green's functions $G^{r(a)}$ are $4\times4$ matrices in Nambu space, 
\begin{equation}
\label{Gr}
G^{r}(x,x',\omega)=
\begin{pmatrix}
G^{r}_{ee}(x,x',\omega)&G^{r}_{eh}(x,x',\omega)\\
G^{r}_{he}(x,x',\omega)&G^{r}_{hh}(x,x',\omega)
\end{pmatrix}\,,
\end{equation}
where the diagonal elements are the regular electron-electron and hole-hole Green's functions, while the off-diagonal corresponds to the anomalous electron-hole part. Here each element is a $2 \times 2$ matrix in spin-space.
The electron-electron part allows us to calculate the LDOS
\begin{equation}
\label{LDOS}
\rho_{}(x,\omega)=-\lim_{x\to x'}\frac{1}{\pi}{\rm Im}{\rm Tr}[G^{r}_{ee}(x,x',\omega)],
\end{equation}
while the electron-hole part provides the pairing amplitudes. The LDOS is generally one of the most experimentally accessible quantities in a system and can be measured directly using, for example, different scanning tunneling probes.\cite{PhysRevLett.77.3025, doi:10.1063/1.1331328}
Notice that the LDOS is proportional to the trace of the imaginary local ($x=x'$) Green's function. Therefore, any connection between LDOS and pairing amplitudes is most naturally occurring for local pair amplitudes, which have $s$-wave symmetry.

Once the retarded and advanced Green's functions are calculated, we can decompose the spin symmetry of the anomalous electron-hole part as 
\begin{equation}
\label{IsolateSpin}
G^{r(a)}_{eh}(x,x',\omega)=f^{r(a)}_{0}(x,x',\omega)\sigma_{0}+\sum_{j=1}^3f^{r(a)}_{j}(x,x',\omega)\sigma_{j}\,,
\end{equation}
where $f_{0}^{r(a)}$ corresponds to spin-singlet, $f_{1,2}^{r(a)}$ to equal spin-triplet and $f_{3}^{r(a)}$ to mixed spin-triplet retarded (advanced) pairing amplitudes.  
These components all necessarily obey Fermi-Dirac statistics and are therefore antisymmetric under the simultaneous exchange of positions and frequency, 
\begin{equation}
\label{antif_main}
\begin{split}
f^{r}_{0}(x,x',\omega)&=f_{0}^{a}(x',x,-\omega)\,,\\
f^{r}_{j}(x,x',\omega)&=-f_{j}^{a}(x',x,-\omega)\,,\\
\end{split}
\end{equation}
with $j=1,2,3$ labeling the spin-triplet pairing amplitudes. Note that when analyzing the symmetry with respect to frequency, the change $\omega\rightarrow-\omega$ requires a pass from the retarded Green's function into the advanced function since each are only defined on parts of the time (frequency) axis, see Appendix\,\ref{AppG} for more details.
From this we can isolate the even- and odd-frequency components as
\begin{equation}
\label{Odd_Even_text}
\begin{split}
f_{i}^{r,\rm{E}}(x,x',\omega)&=\frac{f_{i}^{r}(x,x',\omega)+f_{i}^{a}(x,x',-\omega)}{2}\,,\\
f_{i}^{r,{\rm O}}(x,x',\omega)&=\frac{f_{i}^{r}(x,x',\omega)-f_{i}^{a}(x,x',-\omega)}{2}
\end{split}
\end{equation}
for all spin components ($i = 0,1,2,3$).

Taking the full antisymmetry condition of the anomalous Green's functions into account, which includes spin, spatial, and frequency dependence, we are left with a total of four allowed symmetry classes.\cite{PhysRevLett.98.037003,Eschrig2007,PhysRevLett.99.037005,PhysRevB.76.054522, Nagaosa12} These symmetries are Even-frequency, spin-Singlet, Even in space (ESE), Odd-frequency, spin-Singlet, Odd in space (OSO), Even-frequency, spin-Triplet, Odd in space (ETO), and Odd-frequency, spin-Triplet Even in space (OTE). 
As we will see, NS structures along 2DTI edges generally host all of these four components even though the superconductor itself only has conventional spin-singlet $s$-wave symmetry (ESE). 
However, considering the stability against disorder and also favoring a direct comparison with the  LDOS which is a fully local property (extracted at $x' = x$), we will here primarily be concerned with states with local or equivalently uniform $s$-wave symmetry with $f_{i}(x,x,\omega)\equiv f_{i}(x,\omega)$, which are the ESE and OTE components. 

\section{Pairing amplitudes and LDOS}
\label{sec2}
We now turn to the main objective in this work which is to analyze the pair amplitudes in NS, NSN, and SNS superconducting hybrid junctions at the edge of a 2DTI and its consequences for the LDOS. In normal NS junctions, i.e. not in TIs, it has previously been shown that when a superposition of spin-singlet and singlet-triplet superconductivity is present, the state with larger pairing magnitude dominates the induced superconductivity\cite{PhysRevB.90.085438} and with remarkable consequences for the LDOS.\cite{PhysRevB.87.104513} 
Before going into details, we stress that in this work both the retarded and advanced equal spin-triplet pairing amplitudes are zero since we do not have any magnetic order in our hybrid junctions, i.e.~ $f^{r(a)}_{1,2}(x,x',\omega)=0$. We will therefore refer to the mixed spin-triplet component $f_{3}^{r}$ simply as the spin-triplet component. Moreover, we refer to $f^{r}_{i}$ as the pairing amplitude, while $|f^{r}_{i}|=\sqrt{f_{i}^{r}f_{i}^{r,*}}$ is the pairing magnitude, unless otherwise specified. 

\subsection{NS junction}
We first focus on the simplest situation, which consists of a semi-infinite NS junction, where the N region occupies $x<0$ and S $x>0$.
Before proceeding further we point out that NS junctions at the metallic edge have previously been found, both theoretically and experimentally, to allow perfect (local) Andreev reflection for energies within the superconducting gap with totally suppressed normal reflection processes (backscattering).\cite{PhysRevB.79.161408,PhysRevB.82.081303,PhysRevLett.109.186603}

Extracting the Green's function from the interface scattering processes we find in the normal region N for $x<0$ (see Appendix \ref{NSApp} for details) the even and odd-frequency pairing amplitudes
\begin{equation}
\label{f_NS_N2_Main}
\begin{split}
f^{r,{\rm E}}_{0, {\rm N}}(x,x',\omega)&=\frac{a_{2}(\omega)}{2iv_{F}}\,C_{xx'}\,{\rm e}^{-i(x+x')/\xi_{\omega}},\\
f^{r,{\rm O}}_{0, {\rm  N}}(x,x',\omega)&=-\frac{a_{2}(\omega)}{2v_{F}}\,S_{xx'}\,{\rm e}^{-i(x+x')/\xi_{\omega}},\\
f^{r,{\rm E}}_{3,{\rm  N}}(x,x',\omega)&=\frac{a_{2}(\omega)}{2v_{F}}\,S_{xx'}\,{\rm e}^{-i(x+x')/\xi_{\omega}},\\
f^{r,{\rm O}}_{3,{\rm N}}(x,x',\omega)&=-\frac{a_{2}(\omega)}{2iv_{F}}\,C_{xx'}\,{\rm e}^{-i(x+x')/\xi_{\omega}},\\
\end{split}
\end{equation}
where $C_{xx'}={\rm cos}[k_{\mu}(x-x')]$,  $S_{xx'}={\rm sin}[k_{\mu}(x-x')]$, and $k_{\mu}=\mu/v_{F}$, $\xi_{\omega}=v_{F}/\omega$. Here the superscripts indicate the retarded Green's function (r), the even (E) and odd (O) frequency components, while subscripts stand for spin-component (0 or 3) and normal region (N).
Moreover, $a_{2}$ represents the Andreev reflection amplitude for an incident hole from N at the NS interface. At the metallic edge of a 2DTI $a_2$ it is the same as the Andreev reflection for an incident electron from N  at the NS interface, $a_{1}$. For energies within the superconducting gap, $a_{1,2}= {\rm e}^{-i\eta(\omega)}$, where $\eta(\omega)={\rm arccos}(\omega/\Delta)$.
Despite the non-obvious frequency dependence of  Eqs.~\eqref{f_NS_N2_Main} it is straightforward to verify their symmetry with respect to frequency. 
For instance, the local ESE amplitude $f^{r,E}_{0}(x,\omega) \equiv f^{r,E}_{0}(x,x,\omega)=a_{2}(\omega){\rm e}^{-2ix/\xi_{\omega}}/(2iv_{F})$ and thus reversing the sign of $\omega$ leads to $f^{r,E}_{0}(x,-\omega)\equiv f^{a,E}_{0}(x,-\omega)=-a^{*}_{2}(-\omega){\rm e}^{-2ix/\xi_{\omega}}/(2iv_{F})=f^{r,E}_{0}(x,\omega)$, since $a_2^*(-\omega) = -a_2(\omega)$. This confirms the even frequency dependence of the ESE pairing amplitude. 

The amplitudes in Eqs.\,(\ref{f_NS_N2_Main}) correspond to ESE, OSO, ETO, OTE amplitudes, respectively, and they are all finite for $x\neq x'$ and  finite Andreev reflection amplitude $a_{2}$. 
The emergence of all symmetry classes at interfaces of junctions, as the ones discussed here, is not surprising. Indeed, the junction breaks translational symmetry which allows for mixing between different spatial symmetries. Moreover, the 2DTI also allows for mixing between different spin states, without the presence of a magnetic field. Thus all symmetry classes preserving the antisymmetric Fermi-Dirac statistic will appear at the interface.
Moreover, we observe that the even-frequency (odd-freqeuncy) spin-singlet and odd-frequency (even-frequency) spin-triplet pairing amplitudes are equal to each other in magnitude, namely $|f^{r,{\rm E(O)}}_{0}(x,x',\omega)|=|f^{r,{\rm O(E)}}_{3}(x,x',\omega)|$.
Considering only $s$-wave pairing, this means that ESE and OTE contribute equally to the superconducting state in N. This situation with very strong odd-frequency components is due to the unique nature of the 1D metallic edge.
We also note that Eqs.\,(\ref{f_NS_N2_Main}) suggest that the pairing amplitudes in N do not decay into N, but rather remain finite as a pure consequence of the NS interface. This has to be understood to happen only very close to the NS interface, where Andreev reflection is manifested.

Performing the same analysis as above for the superconducting region S, $x>0$, leads us to naturally distinguish between two different terms in the pairing amplitudes, attributed to the bulk (B) and to the interface (I), with the total pairing amplitude $f = f_{\rm B} + f_{\rm I}$. For energies within the superconducting gap, $|\omega|<\Delta$, we find
\begin{equation}
\label{f_NS_Sx_main}
\begin{split}
f^{r,{\rm E}}_{0,{\rm B}}(x,x',\omega)&=Z(\omega){\rm e}^{-\kappa(\omega)|x-x'|}\,C_{xx'}\,,\\
f^{r,{\rm E}}_{3,{\rm B}}(x,x',\omega)&=iZ(\omega){\rm e}^{-\kappa(\omega)|x-x'|}\,S_{xx'}\,,\\
f^{r,{\rm O}}_{i,{\rm B}}(x,x',\omega)&=0\,
\end{split}
\end{equation}
and \begin{equation}
\label{f_NS_Sy_main}
\begin{split}
f^{r,{\rm E}}_{0,{\rm I}}(x,x',\omega)&=\frac{a_{3}(\omega)}{2iv_{F}}{\rm e}^{-\kappa(\omega)(x+x')}
B(\omega)
C_{xx'}\,,\\
f^{r,{\rm O}}_{0,{\rm I}}(x,x',\omega)&=\frac{a_{3}(\omega)}{2v_{F}}{\rm e}^{-\kappa(\omega)(x+x')}\,S_{xx'}\,,\\
f^{r,{\rm E}}_{3,{\rm I}}(x,x',\omega)&=\frac{a_{3}(\omega)}{2v_{F}}{\rm e}^{-\kappa(\omega)(x+x')}
B(\omega)
S_{xx'}\,,\\
f^{r,{\rm O}}_{3,{\rm I}}(x,x',\omega)&=\frac{a_{3}(\omega)}{2iv_{F}}{\rm e}^{-\kappa(\omega)(x+x')}\,C_{xx'}\,,
\end{split}
\end{equation}
with $Z(\omega)=(1/iv_{F})/[{\rm e}^{i\eta(\omega)}-{\rm e}^{-i\eta(\omega)}]$, $B(\omega)=-i{\rm cot}[\eta(\omega)]$, $\kappa(\omega)=\sqrt{\Delta^{2}-\omega^{2}}$, and $a_3 = -a_1.$
For details on the derivation we refer to Appendix \ref{NSApp}.
The division into bulk and interface properties is evident for energies within the superconducting gap. Here the interface components exhibit an exponential decay\cite{PhysRevB.92.100507} proportional to ${\rm e}^{-\kappa(\omega)(x+x')}$, while the bulk components have an exponential decay proportional to ${\rm e}^{-\kappa(\omega)|x-x'|}$. Thus, locally at $x=x'$, the bulk components are independent of spatial coordinates, confirming their bulk nature, while the interface components display an exponential decay into the bulk of the S region. Notice also that the interface amplitudes are all proportional to the Andreev reflection amplitude $a_3 = -a_{1}$ and thus purely a consequence of the NS interface.
Analyzing Eqs.\,(\ref{f_NS_Sx_main}-\ref{f_NS_Sy_main}) we find that only the even-frequency ESE and ETO components are finite in the bulk. At the interface, however, all symmetry classes (ESE, OSO, ETO, OTE) are present. 
 Even limiting the interest to the disorder robust $s$-wave states leaves both ESE and OTE amplitudes in the interface region. 
 Note that an OTE component does not appear in normal metal NS junctions, but there instead a magnetic field is necessary to generate the OTE component.

\begin{figure}[!ht]
\centering
\includegraphics[width=.45\textwidth]{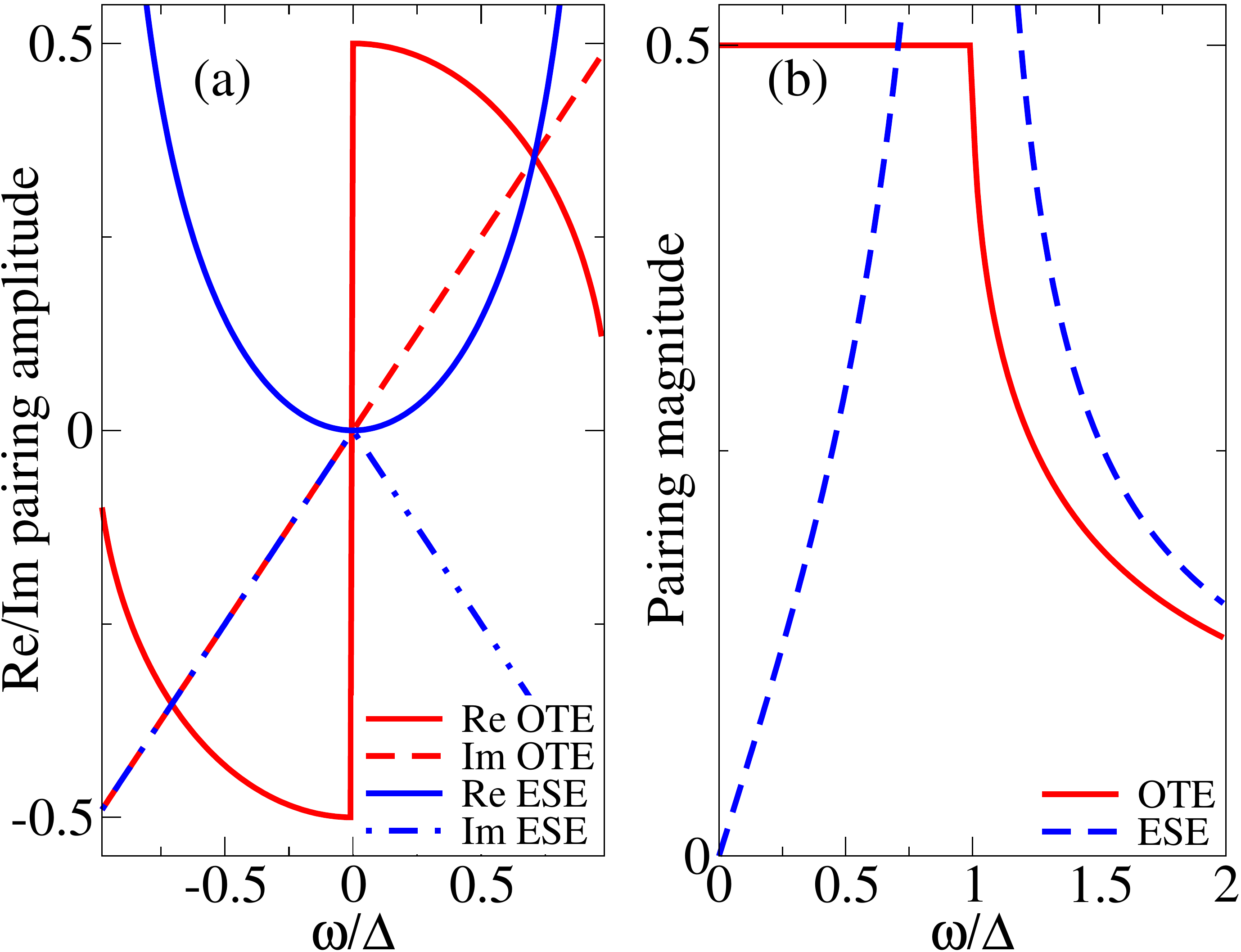} 
\caption{(Color online) Real and imaginary pairing amplitudes (a) and pairing magnitudes (b) of the interface OTE and ESE at $x=x'=0$ as a function of $\omega$. Notice the even and odd behaviors with respect to $\omega$ and how OTE is dominant over the ESE for $|\omega|<\Delta/\sqrt{2}$.}
\label{figNSx}
\end{figure}
In Fig.\,\ref{figNSx} we display the frequency dependence of the pairing amplitudes at the interface,  $x=x'=0$. This confirms the odd frequency dependence of the OTE component, while the ESE is fully even in frequency. While the figure only shows the $x=x'=0$, we have verified the frequency dependence also for $x \neq x'$.
Notice that the real part of OTE exhibits an abrupt transition across zero, which arises due to the discontinuity between $f^{a}$ and $f^{r}$ at zero frequency.
Thus, the magnitude of the OTE component is finite even for $\omega=0$, as is shown in Fig.\,\ref{figNSx}(b). One could argue that at $\omega=0$ the retarded and advanced Green's functions are not well defined and therefore this discussion might lead to a wrong interpretation of the emergence of OTE. We can rule out such a possibility by pointing out that away from zero frequency, but well below $\Delta$, the OTE component is still finite and even clearly dominant over the ESE component. 
In fact, we find that the OTE amplitude dominates over ESE at the interface for $|\omega|<\Delta/\sqrt{2}$.

\begin{figure}[!ht]
\centering
\includegraphics[width=.45\textwidth]{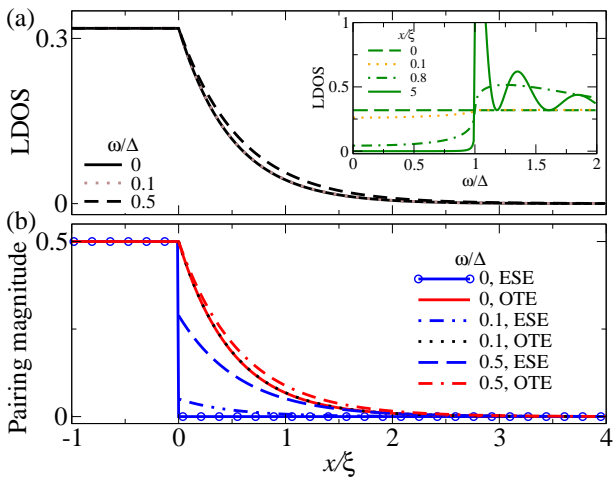} 
\caption{(Color online) Spatial dependence of the LDOS (a) and the normal and interface ESE and OTE pairing magnitudes $|f_{\rm N,I}^{r}|$ (b) as a function of $x=x'$ for an NS junction with the interface at $x =0$.
Inset shows the LDOS as a function of $\omega$. Parameters: $v_{F}=1$, $\Delta=1$.}
\label{figNS}
\end{figure}
In order to establish a connection with the LDOS, we compare the LDOS with the local, $s$-wave, pairing amplitudes at $x=x'$.
In Fig.\,\ref{figNS} we present the spatial dependence of the LDOS and the local pairing magnitudes in the normal and interface regions with the interface placed at $x=0$.
In the normal region, $x<0$, the LDOS acquires a finite value $\rho_{N}(x,\omega) =1/(\pi v_{F})$ independent of both position and energy, as shown in Fig.\,\ref{figNS}(a). This naturally arises because, at the metallic edge of a 2DTI, an incident electron is purely reflected back as a hole without interference.\cite{PhysRev.175.559}
The equal and finite values of ESE $f^{r,{\rm E}}_{0, {\rm N}}(x,\omega)$ and OTE $f^{r,{\rm O}}_{0, {\rm N}}(x,\omega)$ in N, first and last equations in Eqs.\,(\ref{f_NS_N2_Main}), suggest that $\rho_{N}$ arises due to their simultaneous contribution. 

In the superconducting region, $x>0$, the situation is more interesting.  Within the superconducting gap, $|\omega|<\Delta$, the LDOS is 
\begin{equation}
\label{DOS_NS_S_main}
\rho_{\rm S}(x,|\omega|<\Delta)=\frac{1}{\pi v_{F}}{\rm e}^{-2\kappa(\omega)x}\,.
\end{equation}
At the interface, $x=0$, the LDOS is thus finite for energies within the superconducting gap $\Delta$, acquiring its maximum value and then decaying into the bulk of S with a  decay length given by $1/[2\kappa(\omega)]$, and also seen in Fig.\,\ref{figNS}(a). 
For $|\omega|>\Delta$ the inset in Fig.\,\ref{figNS}(a) shows the energy dependence and spatial dependence in the S region. Well within the S region the LDOS is depleted below the gap and shows an oscillatory pattern for $|\omega|>\Delta$, which arises due to interference between the incident and reflected quasiparticles,\cite{PhysRev.175.559} unlike in the N region. 
For the local pairing amplitudes in S we simplify Eqs.\,(\ref{f_NS_Sy_main}) by setting $x = x'$ and obtain
\begin{equation}
\label{eqNSs}
\begin{split}
f^{r,{\rm E}}_{0,{\rm I}}(x,\omega)&=\frac{a_{3}(\omega)}{2iv_{F}}{\rm e}^{-2\kappa(\omega)x}
B(\omega)\,,\\
f^{r,{\rm O}}_{3,{\rm I}}(x,\omega)&=\frac{a_{3}(\omega)}{2iv_{F}}{\rm e}^{-2\kappa(\omega)x}\,,
\end{split}
\end{equation}
for the ESE and OTE interface amplitudes. Notice that the ESE  interface component is multiplied by the factor $B(\omega)$, which leads to important consequences at low energies. 
Remarkably, at the interface and zero energy the ESE interface component is totally suppressed because $B(\omega=0)=0$, while 
the OTE amplitude is still non-zero and even acquires its maximum value, as seen in Fig.\,\ref{figNS}(b). We can therefore directly associate the enhancement of the LDOS close to the interface in S, Fig.\,\ref{figNS}(a), with the OTE pairing and not with the conventional ESE pairing.\footnote{A similar conclusion was achieved in junctions between normal metals and unconventional superconductors,\cite{PhysRevLett.99.037005,PhysRevB.76.054522} which are sensitive to backscattering and disorder unlike our 2DTI system.} Interestingly, this behavior is preserved for energies away from (but close to) zero. For energies close to $\Delta$, however, there is an equal contribution of the ESE component.
We have verified that these results do not change by introducing an insulating barrier at the NS interface, since the Andreev reflection coefficient $a_{1}$ remains unchanged. Thus the 1D metallic edges of 2DTI offers more robust predictions unlike its 2D surface counterparts, which are sensitive to an interface barrier.\cite{PhysRevB.92.205424,PhysRevB.88.075401} We have also only considered a step-like pairing potential $\Delta$. Preliminary results (not shown) using a smoother $\Delta(x)$ across the NS interfaces, present due to inverse proximity effect, show that there is a reduction of the pairing magnitudes at the NS interfaces, characterized by the sharpness of the $\Delta(x)$ profile. However, the classification of the pairing symmetries as well as the main conclusions remain unchanged. This is consistent with the OTE pairing amplitude being generated by an in-surface gradient,\cite{PhysRevB.86.144506} which naturally is maximized for step-edge profiles.

Based on our results we can also relate the conductance across the NS junction to the OTE pairing.
The zero-temperature single mode conductance for an incident electron from N is given by $G_{\rm NS}(\omega)=(e^{2}/h)(1+|a_{1}|^{2}-|b_{1}|^{2})$, where $a_{1}$ and $b_{1}$ are the Andreev and normal reflection amplitudes, respectively. Since backscattering is forbidden in the 2DTI edge $b_{1}=0$. Therefore, the conductance is fully described  by the Andreev reflection coefficient $a_{1}$. At the same time, the interface ESE and OTE pairing amplitudes are both proportional to the Andreev reflection coefficient $a_{3} = -a_1$. At low energies the $B(\omega)$ factor, however, dramatically suppresses the ESE pairing and left is therefore only the OTE contribution.
Thus, any change in the Andreev reflection process is directly manifested in both the conductance across the NS junction and OTE pairing magnitude at low energies.
This allows us to conclude that the main contribution to $G_{\rm NS}$ is given by the OTE component only. Notice that the conductance across NS junctions at the edge of a 2DTI is independent of the chemical potentials in the two regions, unlike the situation with 3DTIs,\cite{0953-8984-27-31-315701} and also becomes constant within the gap while exhibiting a decay outside,\cite{PhysRevB.79.161408}
as is evident from the behavior of $a_{1}$ in Figs.~\ref{figNSx}(b) and also \ref{fig_NSN3}(a). 

To conclude this part we especially stress that the contribution of the OTE pairing to the enhancement of LDOS and conductance across the NS junction is a direct consequence of the metallic edge of 2DTI, where the Andreev reflection is perfect for energies below the gap\cite{PhysRevB.82.081303} and also independent of barrier imperfections at the NS interface.\cite{PhysRevB.92.205424,PhysRevB.88.075401}
Moreover, we point out that Andreev reflection at the edge of a 2DTI has already been experimentally demonstrated\cite{PhysRevLett.109.186603} and significant effort has also been devoted to investigate induced superconductivity at the edge of 2DTIs. \cite{Yacoby14,vlad15,Bocquillon17} Our findings can therefore  help to elucidate the nature of the superconducting pairing in both LDOS and conductance measurements in NS junctions.

\subsection{NSN junction}
Having investigated the simplest NS junction we now turn to the situation with a finite S region. 
We consider a NSN junction at the edge of a 2DTI, where the S region is placed within $0<x<L_{\rm S}$, surrounded by N regions on each side.
We arrive at the pairing amplitudes and the LDOS following the same procedure as above, constructing the retarded Green's function from the allowed scattering processes, see Appendix \ref{NSNApp} for details.

In the two normal regions we find the pairing amplitudes proportional to the Andreev reflection amplitude $a_{2}(\omega,L_{\rm S})={\rm sin}[i\kappa(\omega)L_{\rm S}]/{\rm sin}[i\kappa(\omega)L_{\rm S}-\eta(\omega)]$ but otherwise exactly following the same form as for the pairing amplitudes in the N region of NS junctions given in Eqs.\,(\ref{f_NS_N2_Main}). Thus the only change for the N pairing amplitudes compared to the NS junction is $a_{2}(\omega)\rightarrow a_{2}(\omega,L_{\rm S})$. This is fully consistent with the results in N being purely due to the interfaces. 
In particular, notice that the dependence on $L_{\rm S}$  is only present through $a_{2}(\omega,L_{{\rm S}})$. The magnitude $|a_{2}(\omega,L_{{\rm S}})|$ as a function of $\omega$ for different $L_{{\rm S}}$ is plotted in Fig.\,\ref{fig_NSN3}(a). For $|\omega|<\Delta$ the magnitude increases as the length of the S region increases and saturates to the value found in NS junctions $|a_{2}(\omega)|=|a_{2}(\omega,L_{{\rm S}}\rightarrow\infty)|$ for $L_{{\rm S}}> \xi$, where $\xi=v_{\rm F}/\Delta$ is the superconducting coherence length. For energies above $\Delta$, $|a_{2}(\omega,L_{{\rm S}})|$ develops an oscillatory decay whose amplitude for $\xi<L_{\rm S}\ll\infty$, is higher than that in NS junctions. 

Turning to the middle superconducting region and concentrating on energies below the gap $\Delta$, we obtain the following even and odd-frequency pairing amplitudes divided into bulk contributions
\begin{equation}
\label{f_NSN_S3_main1}
\begin{split}
f_{0,{\rm B}}^{r,{\rm E}}(x,x',\omega)&=2C_{xx'}\Big[\beta_{2}{\rm e}^{-\kappa(\omega)|x-x'|}+\beta_{3}{\rm e}^{\kappa(\omega)|x-x'|}\Big]\,,\\
f_{3, {\rm B}}^{r,{\rm E}}(x,x',\omega)&=2iS_{xx'}\Big[\beta_{2}{\rm e}^{-\kappa(\omega)|x-x'|}+\beta_{3}{\rm e}^{\kappa(\omega)|x-x'|}\Big]\,,\\
f_{0,3,{\rm B}}^{r,{\rm O}}(x,x',\omega)&=0
\end{split}
\end{equation}
and interface contributions
\begin{equation}
\label{f_NSN_S3_main2}
\begin{split}
f_{0,{\rm I}}^{r,{\rm E}}(x,x',\omega)&=C_{xx'}
\Big[\beta_{42}^{-}{\rm e}^{-\kappa(\omega)(x+x')}+\beta_{13}^{-}{\rm e}^{\kappa(\omega)(x+x')}\Big]\,,\\
f_{0,{\rm I}}^{r,{\rm O}}(x,x',\omega)&=-iS_{xx'}
\Big[\beta_{42}^{+}{\rm e}^{-\kappa(\omega)(x+x')}+\beta_{13}^{+}{\rm e}^{\kappa(\omega)(x+x')}\Big]\,,\\
f_{3,{\rm I}}^{r,{\rm E}}(x,x',\omega)&=iS_{xx'}
\Big[\beta_{42}^{-}{\rm e}^{-\kappa(\omega)(x+x')}+\beta_{13}^{-}{\rm e}^{\kappa(\omega)(x+x')}\Big]\,,\\
f_{3,{\rm I}}^{r,{\rm O}}(x,x',\omega)&=-C_{xx'}
\Big[\beta_{42}^{+}{\rm e}^{-\kappa(\omega)(x+x')}+\beta_{13}^{+}{\rm e}^{\kappa(\omega)(x+x')}\Big]\,,\\
\end{split}
\end{equation}
where $\beta_{42(13)}^{\pm}=\beta_{4(1)}\pm\beta_{2(3)}$, $\beta_{4,3}=-{\rm e}^{-2i\eta(\omega)}\beta_{2,1}$, $\beta_{1}=-iZ(\omega){\rm e}^{i\eta(\omega)-\kappa(\omega)L_{\rm S}}/4{\rm sin}[i\kappa(\omega)L_{\rm S}-\eta(\omega)]$ and $\beta_{2}=-\beta_{1}{\rm e}^{2\kappa(\omega)L_{\rm S}}$.
First of all, we stress that these results are in agreement with the expressions for the NS junction. Indeed, we find that in the bulk only even-frequency pairing amplitudes exist, while all pairing symmetries have finite contributions at the interface. 
Observe also that the interface components develop an exponential decay from both interfaces proportional to ${\rm e}^{\pm \kappa(\omega)(x+x')}$, while the bulk components are proportional to ${\rm e}^{\pm \kappa(\omega)|x-x'|}$ and thus become independent of space for $x=x'$. 

\begin{figure}[htb]
\centering
\includegraphics[width=.45\textwidth]{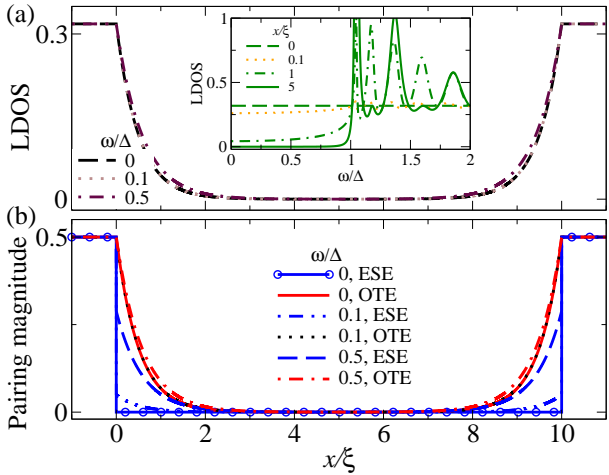} 
\caption{(Color online) Spatial dependence of the LDOS (a) and normal and interface ESE and OTE pairing magnitudes $|f_{{\rm N,I}}^{r}|$ (b) as a function of $x = x'$ in a NSN junction with the interfaces at $x =0, 10\xi$. 
Inset shows the LDOS as a function of $\omega$. 
Parameters: $v_{F}=1$, $\Delta = 1$, $L_{{\rm S}}=10\xi$.}
\label{fig_f_NSN}
\end{figure}

\begin{figure}[!ht]
\centering
\includegraphics[width=.49\textwidth]{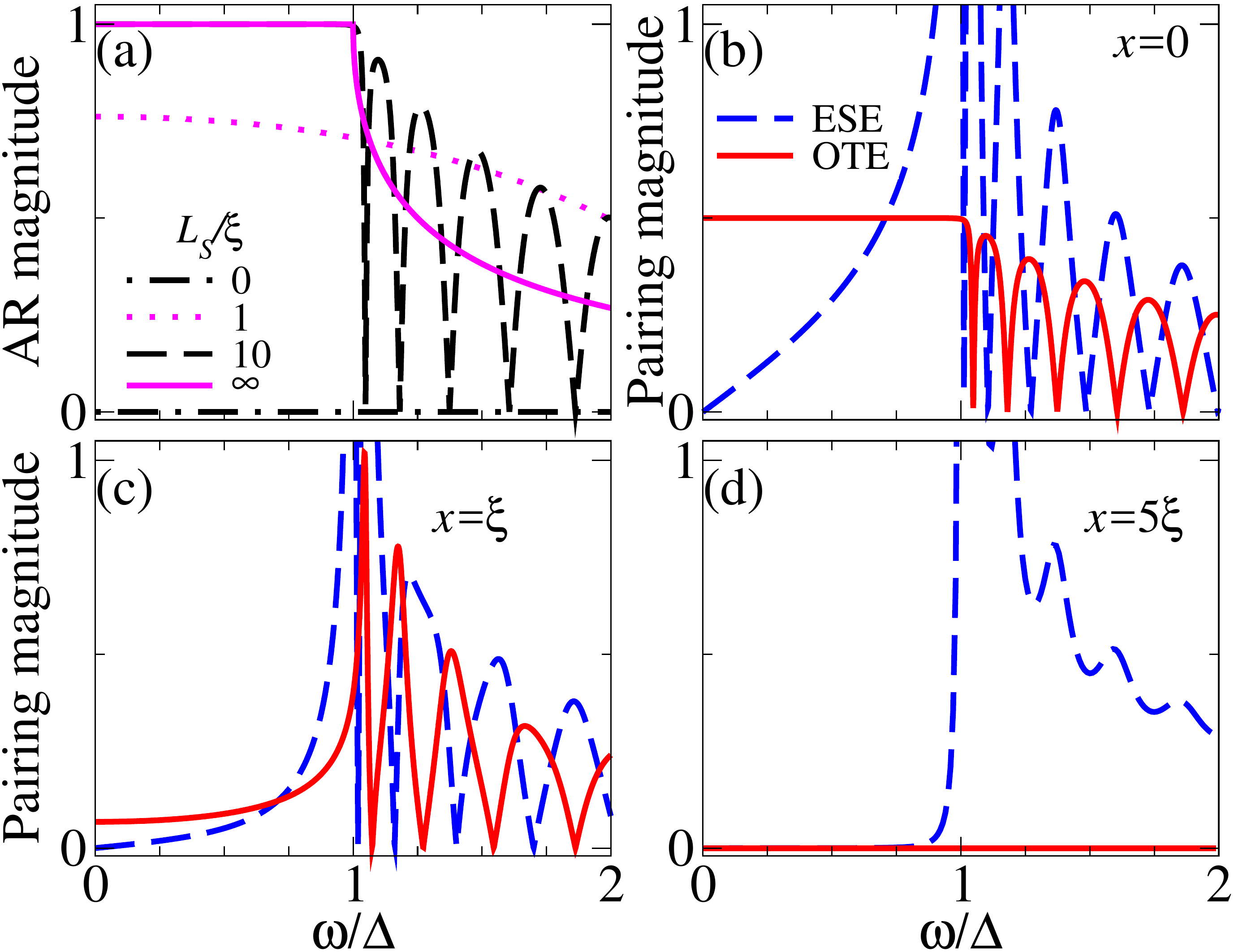} 
\caption{(Color online) Andreev reflection (a) and ESE and OTE interface pairing magnitudes $|f_{0,3,{\rm I}}^{r}|$ as a function of $\omega$ in a NSN junction at the left interface $x=0$ (b), $x=\xi$ (c), and in the middle of S $x=5\xi$ (d).  Parameters: $v_{F}=1$, $\Delta = 1$, $L_{{\rm S}}=10\xi$.}
\label{fig_NSN3}
\end{figure}

Next we proceed to analyze how the local pairing amplitudes for $x=x'$ are related to the LDOS.
The regular Green's functions in the N region are exactly the same as in the NS junction case and the LDOS in this N region is subsequently also the same, $\rho_{\rm  N} = 1/(\pi v_{F})$. On the other hand, in the S region we find a much more complex expression for the LDOS (see Appendix \ref{NSNApp} for details). However, it acquires a simple expression at zero energy
 \begin{equation}
\label{LDOS_NSN3x}
\rho_{\rm S}(x,\omega=0)=\frac{1}{\pi v_{F}}\frac{{\rm e}^{-2L_{\rm S}/\xi}{\rm e}^{2x/\xi}+{\rm e}^{-2x/\xi}}{1+{\rm e}^{-2L_{\rm S}/\xi}}\,.
\end{equation}
In Fig.\,\ref{fig_f_NSN}(a) we plot the spatial and frequency dependence of the LDOS. At the interfaces, $x=0, 10\xi$, the LDOS is finite and exhibits its maximum value $1/(\pi v_{F})$. It then exponentially decays from both interfaces into the S region with a decay length for $\omega=0$ given by $\xi/2$. This behavior is preserved even for energies away from zero as can be observed in Fig.\,\ref{fig_f_NSN}(a). For $\omega\neq0$ the decay length is given by $1/[2\kappa(\omega)]$.
We compare the LDOS with the local pairing amplitudes at $x = x'$, derived from Eqs.\,(\ref{f_NSN_S3_main1}) and (\ref{f_NSN_S3_main2}), giving the finite components
\begin{equation}
\label{nsn2}
\begin{split}
f_{0,{\rm B}}^{r,{\rm E}}(x,\omega)&=\beta_{2}-\beta_{1}{\rm e}^{-2i\eta(\omega)}\,,\\
f_{0,{\rm I}}^{r,{\rm E}}(x,\omega)&=
B_{1}(\omega)
\Big[\beta_{1}{\rm e}^{2\kappa(\omega)x} -\beta_{2}{\rm e}^{-2\kappa(\omega)x}\Big]\,,\\
f_{3,{\rm I}}^{r,{\rm O}}(x,\omega)&=B_{2}(\omega)
\Big[\beta_{1}{\rm e}^{2\kappa(\omega)x} +\beta_{2}{\rm e}^{-2\kappa(\omega)x}\Big]\,,
\end{split}
\end{equation}
which correspond to the ESE class in the bulk and the ESE and OTE classes at the interface, where $B_{1,2}(\omega)=\pm(1\pm{\rm e}^{-2i\eta(\omega)})/2$.
Figure\,\ref{fig_f_NSN}(b) shows the spatial dependence of the interface and normal region pairing magnitudes for different values of $\omega$.
The pairing amplitudes decay from the interfaces at $x=0$ and $x=L_{{\rm S}}$ into the bulk of S with a decay length given by $1/[2\kappa(\omega)]$. Remarkably, this spatial decay is also observed in the LDOS in Fig.\,\ref{fig_f_NSN}(a), directly supporting their relationship.
Interestingly, at zero energy the coefficients of the interface pairing amplitudes acquire different prefactors, $B_{1}(\omega=0)=0$ and $B_{2}(\omega=0)=2$, which gives rise to purely OTE pairing at the interface. A strongly dominating OTE component is preserved even for energies away from $\omega=0$ but well below $\Delta$ as clearly seen in Fig.\,\ref{fig_f_NSN}(b). 

The previous discussion can be clarified by analyzing the frequency dependence of $|f_{{\rm I}}^{r}|$, plotted in Fig.\,\ref{fig_NSN3}(b-d) for different positions in S. 
At the left interface, $x=0$, Fig.\,\ref{fig_NSN3}(b) shows the OTE pairing being finite and remaining constant as $\omega$ increases within the gap. On the other hand, the ESE pairing is zero at $\omega=0$ and increases only following a V-dependence  as $\omega$ increases. 
To derive the condition for the parameter space where OTE pairing dominates over the ESE pairing, we solve $|f_{0,{\rm I}}^{r,{\rm E}}(\omega,x)|=|f_{3,{\rm I}}^{r,{\rm O}}(\omega,x)|$ for $\omega$. As for NS junctions, we find a simple expression at $x=0$, where the OTE component is larger for $|\omega|<\Delta/\sqrt{2}$.
We also point out that the Andreev reflection amplitude $a_1(\omega,L_{\rm S}) = a_{2}(\omega,L_{\rm S})$ is directly connected to the pairing amplitudes. To visualize this statement we compare the Andreev reflection and pairing magnitudes at the the left interface $x=0$ in Figs.\,\ref{fig_NSN3}(a) and (b). For energies within $\Delta$ the Andreev reflection magnitude is finite and constant for finite $L_{\rm S}\gtrsim \xi$. In these junctions it is only the OTE amplitude that is finite and constant, while the ESE amplitude is zero at zero energy, increases linearly with $\omega$, and is subdominant all the way until $\omega = \Delta/\sqrt{2}$. 
As we go inside the S region, at $x=\xi$, the OTE pairing magnitude is reduced and the ESE pairing enhanced, both taking a U-shape form, see Fig.\,\ref{fig_NSN3}(c). Finally in the middle of the S region  at $x = 5\xi$, the OTE magnitude is totally suppressed, while a BCS-like gap is fully developed for the ESE component, see Fig.\,\ref{fig_NSN3}(d). This is also what we expect due to the ESE nature of the induced superconducting pairing in the bulk. Notice that in the inset of Fig.\,\ref{fig_f_NSN}(a) we see how the LDOS at low energies is finite at the interface, $x=0$, exhibits the same U-shape observed in the pairing magnitudes at $x=\xi$, while deep in the S bulk at $x=5\xi$ the expected BCS-like induced gap is totally developed. 
Thus the LDOS directly follows the magnitude of the OTE  component $|f_{3,{\rm I}}^{r,{\rm O}}|$ and we conclude that the enhancement of the LDOS in S close to the interfaces arises mainly due to the OTE contribution. This relationship is exact at $\omega=0$, as the pairing there is purely OTE. For $0<\omega<\Delta/\sqrt{2}$, when the ESE pairing magnitude is finite, the OTE magnitude is still dominating close to the interface and thus has the largest influence on the LDOS.
This confirms that the emergence of the exotic odd-frequency superconductivity is intrinsically contained in the LDOS at the interface of NSN junctions at the edge of 2DTI, which supports our findings for NS junctions.

\subsection{SNS junction}
We finally turn to a SNS junction along the edge of a 2DTI, where the finite normal region with $\Delta=0$ is restricted to $0<x<L_{\rm N}$. We further set the order parameters in  the left and right S region to $\Delta_{{\rm L}}=\Delta$ and $\Delta_{{\rm R}}=\Delta\,{\rm e}^{i\phi}$, respectively. We arrive at the expressions for the LDOS and all pairing amplitudes following the same formalism as used for NS and NSN junctions, see Appendix \ref{SNSApp} for a detailed derivation.

As for NS and NSN junctions, all pairing amplitude symmetry classes are also present in SNS junctions. Here we however also obtain a dependence on the superconducting phase across the junction. As we will see, this phase dependence gives rise to a very strong connection between the LDOS and the local $s$-wave pairing amplitudes.
Let us first focus on the normal N region. We find the even and odd-frequency components of the pairing amplitudes given by
\begin{equation}
\label{SNS_N_main}
\begin{split}
f_{0,{\rm N}}^{r,{\rm E}}(x,x',\omega)&=W_{+}(\omega){\rm cos}[k_{\mu}(x-x')]\,,\\
f_{0,{\rm N}}^{r,{\rm O}}(x,x',\omega)&=W_{-}(\omega)i{\rm sin}[k_{\mu}(x-x')]\,,\\
f_{3,{\rm N}}^{r,{\rm E}}(x,x',\omega)&=W_{+}(\omega)i{\rm sin}[k_{\mu}(x-x')]\,,\\
f_{3,{\rm N}}^{r,{\rm O}}(x,x',\omega)&=W_{-}(\omega){\rm cos}[k_{\mu}(x-x')]
\end{split}
\end{equation}
where $W_{\pm}(\omega)=[m_{5}{\rm e}^{i(x+x')/\xi_{\omega}}\pm m_{6}{\rm e}^{-i(x+x')/\xi_{\omega}} ]/2$. 
The coefficients $m_{5}=p_{1}\tilde{r}_{4}\alpha_{1}$ and 
$m_{6}=q_{2}\tilde{s}_{3}\alpha_{4}$ contain important information about the scattering processes across 
the junction. 
In fact, $p_{1}$ is the amplitude for electron transmission from left S into N, 
$\tilde{r}_{4}$ gives the Andreev reflection at the left SN interface of a left-moving electron into a hole, 
$q_{2}$ gives the Andreev reflection at the right SN interface of a right-moving hole into an electron, and
$\tilde{s}_{3}$ is the hole transmission from right S into N.
Moreover, we find 
$m_{6}(\omega,L_{\rm N},\phi)=m_{5}(\omega,L_{\rm N},-\phi){\rm e}^{i(2L_{\rm N}/\xi_{\omega}+\phi)}$ 
and for energies within the gap, $|\omega|<\Delta$,
\begin{equation}
\label{m5}
\begin{split}
m_{5}(\omega,L_{\rm N},\phi)&=-\frac{1}{2v_{F}}\frac{{\rm e}^{i(\phi/2- L_{\rm N}/\xi_{\omega})}}{{\rm sin}[\eta(\omega)-L_{\rm N}/\xi_{\omega}+\phi/2]}\,.
\end{split}
\end{equation}
The first general observations are that the pairing amplitudes, given by Eqs.\,(\ref{SNS_N_main}), become dependent on the superconducting phase difference $\phi$ through the coefficients $m_{5,6}$. Note also that all symmetry classes have finite amplitudes. By simple inspection we notice that they do not exhibit any decay but, for frequencies away from zero, rather develop an oscillatory behavior which extends over the whole N region. 

For the two S regions we obtain similar expressions for the pairing amplitudes as for S in the NS junction case. In fact, they include two components that represent the contributions from the interface, proportional to ${\rm e}^{\pm \kappa(\omega)(x+x')}$, 
and two components associated with the bulk and thus proportional to ${\rm e}^{-\kappa(\omega)|x-x'|}$. Here the $\pm$ signs indicate that the exponential decay occurs from both interfaces and into the bulk of the left and right S regions, respectively. Exact expressions are given in Appendix \ref{SNSApp}, but they are not needed here for the future discussion.

Let us now focus on the local pairing amplitudes ESE and OTE present at $x = x'$ and their 
relation to the LDOS. We have verified that the results found in N and in the left and right S regions
provide the same information and thus for clarity we only discuss the results in the N region. 
The N region components, given by Eqs.\,(\ref{SNS_N_main}), for energies within the gap $|\omega|<\Delta$ at $x =x'$ reduce to
\begin{equation}
\label{SNS_N2}
\begin{split}
f_{0,{\rm N}}^{r,{\rm E}}(x,\omega)&=\frac{1}{2}\Big[m_{5}{\rm e}^{2i x/\xi_{\omega}}+m_{6}{\rm e}^{-2ix/\xi_{\omega}} \Big]\,,\\
f_{3,{\rm N}}^{r,{\rm O}}(x,\omega)&=\frac{1}{2}\Big[m_{5}{\rm e}^{2ix/\xi_{\omega}}-m_{6}{\rm e}^{-2ix/\xi_{\omega}} \Big]\,.
\end{split}
\end{equation}
These expressions for the pairing amplitudes can be shown to be directly connected to the LDOS and especially its phase dependence.
By analyzing $m_{5,6}$ we notice that the zeros of their denominators appear at
\begin{equation}
\label{ABS}
2\eta(\omega)-2L_{\rm N}/\xi_{\omega}\pm\phi=2\pi n\,.
\end{equation} 
But this equation is also the exact quantization condition of the Andreev bound states (ABSs) in SNS junctions at the edge of a 2DTI with a normal region of length $L_{\rm N}$,\cite{PhysRevLett.100.096407,PhysRevB.86.214515} see also Appendix \ref{SNSApp}.  These ABSs naturally give very strong peaks in the subgap LDOS in the N region, but here we, quite remarkably, instead obtained them from the pairing amplitudes given by Eqs.\,(\ref{SNS_N2}). This immediately establish a direct relationship between local pairing amplitudes and the LDOS.
A short SNS junction, $L_{\rm N}\ll\xi$, only hosts one pair of ABSs at energies $\omega_{\pm}(\phi)=\pm \Delta {\rm cos}(\phi/2)$, which are $4\pi$-periodic and develop zero-energy crossings at $\phi=\pi(2n-1)$, $n=1,2,\ldots$ protected by time-reversal symmetry.\cite{PhysRevLett.100.096407,PhysRevB.79.161408,PhysRevB.86.214515,PSSB:PSSB201248385,PhysRevB.88.075401} 
These zero-energy protected crossings give rise to a bound state with Majorana-like properties.\cite{PhysRevB.79.161408} In a long SNS junction, $L_{\rm N}\gg\xi$, on the other hand, many more ABSs fit within the superconducting energy gap. Here the ABSs appear at  $\omega_{\pm,n}(\phi)=(v_{F}/2L)[2\pi(n+1/2)\pm \phi]$, which gives a linear dispersion with respect to the superconducting phase difference $\phi$.

\subsubsection{Short junctions}
\begin{figure}[!ht]
\centering
\includegraphics[width=.49\textwidth]{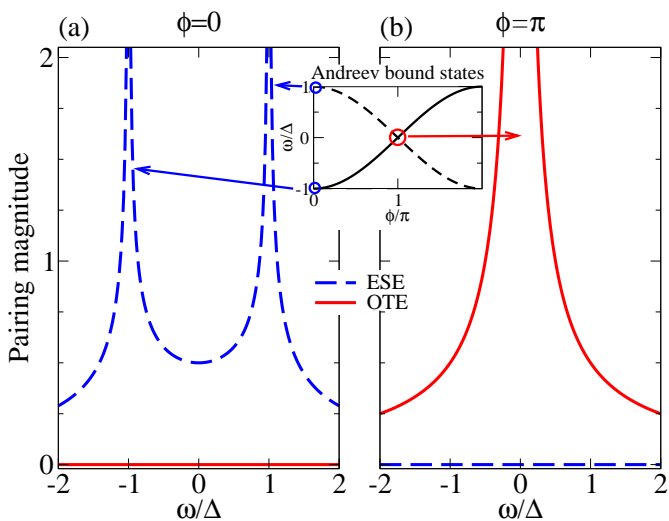} 
\caption{(Color online) ESE and OTE pairing magnitudes in the normal region for a vanishingly short SNS junction as a function of $\omega$ at $\phi=0$ (a) and $\phi=\pi$ (b). Inset shows the ABS spectrum as a function of $\phi$. Parameters: $v_{F}=1$, $\Delta = 1$, $L_{\rm N}=0$, $x=0$.}
\label{fig_SNS_N}
\end{figure}

To further analyze the connection between pairing amplitudes and ABSs energies we first investigate the situation of a short junction, $L_{\rm N}\ll\xi$, corresponding to the limit $L_{\rm N}\rightarrow 0$. 
From Eq.\,(\ref{m5}) we obtain  $m_{5}(\omega,L_{\rm N}=0,\phi)=-(1/2v_{\rm F}){\rm e}^{i\phi/2}/{\rm sin}[\eta(\omega)+\phi/2]$ and $m_{6}(\omega,L_{\rm N}=0,\phi)=m_{5}(\omega,L_{\rm N}=0,-\phi){\rm e}^{i\phi}$. By simple inspection we notice that this simplified dependence on $\phi$ has profound consequences for the pairing amplitudes in Eqs.\,(\ref{SNS_N2}). 
At $\phi=2\pi n$, for $n=0,1,2,\ldots$, the coefficients are the same, i.e.,~$m_{5}(\omega,L_{\rm N}=0,2\pi n)=m_{6}(\omega,L_{\rm N}=0,2\pi n)$, and therefore
\begin{equation}
\begin{split}
f_{0,{\rm N}}^{r,{\rm E}}(x=0,\omega)&=-\frac{1}{2v_{F}}\frac{1}{{\rm sin}[\eta(\omega)]}\,,\\
f_{3,{\rm N}}^{r,{\rm O}}(x=0,\omega)&=0\,.
\end{split}
\end{equation} 
Thus only ESE pairing survives, while the OTE term is completely vanishing. 
This can be understood from the meaning of coefficients $m_{5,6}$. Although they correspond to different (electron and hole) scattering processes, at zero phase difference, they are equal and destructively interfere, which gives rise to zero OTE  but finite ESE pairing.
To visualize these results, we plot in Fig.\,\ref{fig_SNS_N}(a) the ESE and OTE pairing magnitudes as a function of $\omega$ at $\phi=0$. Observe that the OTE magnitude is completely suppressed, while the ESE contribution exhibits resonant peaks at the gap edges $\pm \Delta$. These ESE peaks are in full agreement with the ABSs energies, $\omega_{\pm}(2\pi n)=\pm \Delta$ for  $n=0,1,2,\ldots$, which at $\phi = 0$ merges with the continuum (indicated by blue arrows).

At $\phi=\pi(2n-1)$ for $n=1,2,\ldots$, the coefficients instead acquire opposite values, $m_{5}(\omega,L_{\rm N}=0,\pi(2n-1))=-m_{6}(\omega,L_{\rm N}=0,\pi(2n-1))$, which again affects the pairing amplitudes in a crucial way
\begin{equation}
\begin{split}
f_{0,{\rm N}}^{r,{\rm E}}(x=0,\omega)&=0\,,\\
f_{0,{\rm N}}^{r,{\rm O}}(x=0,\omega)&=-\frac{i}{2v_{F}}\frac{1}{{\rm cos}[\eta(\omega)]}\,.
\end{split}
\end{equation}
Here the ESE contribution is instead completely suppressed, while the OTE term acquires a finite value, as is also shown in Fig.\,\ref{fig_SNS_N}(b).  
At $\phi=\pi(2n-1)$ the ABSs cross zero energy, $\omega_{\pm}(\pi(2n-1))=0$ for $n=1,2,\ldots$, developing a protected crossing and also introducing a resonant peak at $\omega=0$  in the LDOS.
Other work has found similar phase dependence for a combination of spin-orbit coupling and magnetism.\cite{PhysRevB.92.024501} Very importantly, in the 2DTI we obtain that no magnetism is necessary.
This LDOS peak (indicated by red arrow) at zero energy is directly reflected in the OTE pairing amplitude also peaking at zero energy.\footnote{A zero energy peak in the odd-frequency pairing was also reported in junctions with unconventional superconductors,\cite{PhysRevLett.74.3451,PhysRevB.76.054522} where the pairing function is peaked due to the sign change of the pairing potential. Moreover, a zero energy peak in the OTE amplitude was also predicted in Rashba double wires,\cite{Ebisu16} where Rashba coupling and the additional wire index allows for OTE pairing.}
The mechanism is here the same as mentioned before: two scattering processes, Andreev reflection and normal transmission for electrons and holes, become exactly opposite at $\phi=\pi$, unlike  at $\phi=0$ where they are the same. Therefore it exists a clear interference pattern with distinguishable  signatures for each process. We stress that both results, the total suppression of OTE and ESE at $\phi=0$ and $\phi=\pi$, respectively, arise due to the metallic nature of the 2DTI edge. In normal metals finite backscattering processes usually induce finite normal reflection amplitudes.

Before concluding we would like to also point out the connection between OTE and supercurrents $I(\phi)$. In both short and long junctions the supercurrent  can be calculated from the energy levels $\omega_{n}$  as\cite{Beenakker:92,zagoskin,Cayao17b} $I(\phi)=-(2e/\hbar) \sum_{n>0}d\omega_{n}(\phi)/d\phi$. In short junctions only the pair of ABSs within the gap contribute to $I(\phi)$. Thus, for short junctions with the ABSs given by $\omega_{\pm}(\phi)=\pm \Delta {\rm cos}(\phi/2)$, the individual current contribution becomes $I_{\pm}(\phi)= \pm I_{\rm c} {\rm sin}(\phi/2)$, where $I_{\rm c}=e\Delta/\hbar$ is the maximum supercurrent across the junction. Provided fermion parity conservation $I(\phi)$  notably exhibits the well-known $4\pi$-periodic fractional Josephson effect\cite{PhysRevB.79.161408} in $\phi$. Notice that $I_{\pm}(\phi)$ is  zero at $\phi=0$ while at $\phi=\pi$ it acquires  its maximum value. The former arises because there are no ABSs in the junction, while in the latter case the ABSs develop a protected crossing at zero-energy, giving rise to the resonant peak in the LDOS. This resonant peak is, as shown above, purely due to OTE pairing. Thus, the maximum supercurrent, i.e.~the critical current $I_{\rm c}$, exhibits contributions from purely OTE pairing, since at $\phi=\pi$ only OTE pairing is present.

We can thus conclude that the pairing amplitudes entirely capture the LDOS through the emergence of ABSs, and, remarkably, the zero energy peak in the LDOS is purely a consequence of the OTE pairing as shown in Fig.\,\ref{fig_SNS_N}(b). These findings are in agreement with the odd-frequency nature previously found for a single Majorana state in time-reversal symmetry breaking systems.\cite{PhysRevLett.70.2960,Nagaosa12,PhysRevB.87.104513,PhysRevB.92.014513,PhysRevB.92.121404,lutchyn16} However, note that in our setup time-reversal symmetry is preserved and the Majorana mode is therefore necessarily two-fold degenerate. 

\subsubsection{Long junctions}
Next we proceed by increasing the normal region length $L_{{\rm N}}$. Then the number of energy levels, bound in the junction within the energy gap $\Delta$, increases and is proportional to $L_{{\rm N}}/\xi$. In this situation the ESE and OTE pairing amplitudes, given by Eqs.\,(\ref{SNS_N2}), strongly depends on $x$. In what follows we primarily only need to consider two  situations of this spatial dependence, at the left interface, $x=0$, and in the middle of N, $x=L_{\rm N}/2$.

At the left interface, $x=0$, the pairing amplitudes for $\phi=0$ read 
\begin{equation}
\begin{split}
f_{0,{\rm N}}^{r,{\rm E}}(x=0,\omega)&=-\frac{1}{2v_{F}}\frac{{\rm cos}(\omega L_{\rm N})}{{\rm sin}[\eta(\omega)-L_{\rm N}\omega]}\,,\\
f_{3,{\rm N}}^{r,{\rm O}}(x=0,\omega)&=\frac{i}{2v_{F}}\frac{{\rm sin}(\omega L_{\rm N})}{{\rm sin}[\eta(\omega)-L_{\rm N}\omega]}
\end{split}
\end{equation}
and for $\phi=\pi$
\begin{equation}
\begin{split}
f_{0,{\rm N}}^{r,{\rm E}}(x=0,\omega)&=-\frac{1}{2v_{F}}\frac{{\rm sin}(\omega L_{\rm N})}{{\rm cos}[\eta(\omega)-L_{\rm N}\omega]}\,,\\
f_{3,{\rm N}}^{r,{\rm O}}(x=0,\omega)&=\frac{1}{2iv_{F}}\frac{{\rm cos}(\omega L_{\rm N})}{{\rm cos}[\eta(\omega)-L_{\rm N}\omega]}\,.
\end{split}
\end{equation}
\begin{figure}[!ht]
\centering
\includegraphics[width=.49\textwidth]{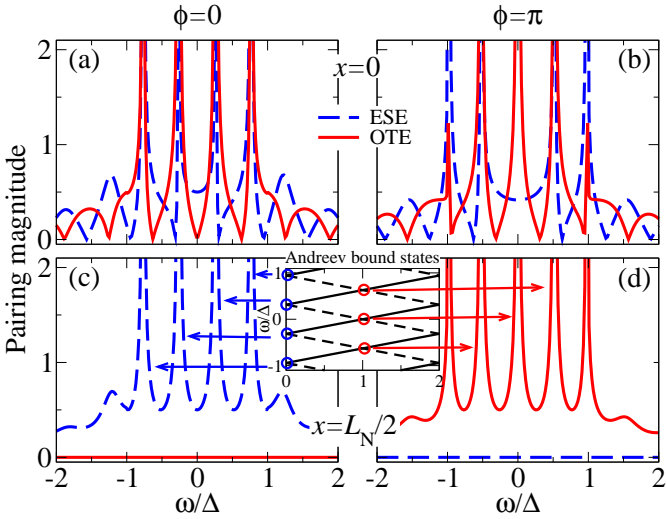} 
\caption{(Color online) ESE and OTE pairing magnitudes in the normal region of a SNS junction as a function of $\omega$ at the left interface $x=0$ (a,b) and in the middle of N $x=L/2$ (c,d) for $\phi=0$ (a,c) and $\phi=\pi$ (b,d). Inset shows the ABS spectrum as a function of $\phi$. Parameters: $v_{F}=1$, $\Delta=1$, $L_{\rm N}=5\xi$.}
\label{fig_SNS2}
\end{figure}
The magnitude of these ESE and OTE pairing terms is plotted in Fig.\,\ref{fig_SNS2}(a,b) at $\phi=0$ and $\phi=\pi$, respectively. 
At $\phi=0$ the ESE and OTE magnitudes reach zero value when $\omega L_{\rm N}=(2n+1)\pi/2$ and $\omega L_{\rm N}=\pi n$ for $n=0,1,2,\ldots$, respectively. In fact, the zeros of ESE correspond to the maxima of OTE and vice-versa.
Specifically, for $\omega=0$ the OTE pairing magnitude is vanishing, while the ESE magnitude remains finite and proportional to $1/(2v_{F})$. 
On the other hand, at $\phi=\pi$ the position of the zeros of the ESE and OTE magnitudes is reversed.
 Remarkably, the OTE magnitude now has a resonant peak at zero energy, while the ESE magnitude remains finite at, again, approximately $1/(2v_{F})$. 
Thus, although the OTE pairing magnitude is dominant around zero energy, the finite value of ESE obscures the pure OTE contribution to the LDOS we found in the short junction limit. 

The coexistence of ESE and OTE pairing in long junctions at the interface $x=0$ can be avoided by probing the pairing amplitudes in the middle of the N region at $x=L/2$, as presented in Fig.\,\ref{fig_SNS2}(c,d). In this case, Eqs.\,(\ref{SNS_N2}) reduce to
\begin{equation}
\label{SNS_N2x}
\begin{split}
f_{0,{\rm N}}^{r,{\rm E}}(x=L_{\rm N}/2,\omega)&=
\bar{m}_{5}+\bar{m}_{6} \,,\\
f_{3,{\rm N}}^{r,{\rm O}}(x=L_{\rm N}/2,\omega)&=
\bar{m}_{5}-\bar{m}_{6} \,,
\end{split}
\end{equation}
where $\bar{m}_{5}(\omega,L_{\rm N},\phi)=(-1/2v_{F}){\rm e}^{i\phi/2}/{\rm sin}[\eta(\omega)-L_{\rm N}/\xi_{\omega}+\phi/2]$ and $\bar{m}_{6}(\omega,L_{\rm N},\phi)=\bar{m}_{5}(\omega,L_{\rm N},-\phi){\rm e}^{i\phi}$. Thus, in the middle of the N region the coefficients $\bar{m}_{5,6}$ lose their length dependence in the numerator, unlike $m_{5,6}$. Moreover, Eqs.\,(\ref{SNS_N2x}) very much resemble the behavior of the pairing amplitudes in short junctions discussed before.
From Eqs.~\eqref{SNS_N2x} it is simple to conclude that at $\phi=0$, $\bar{m}_{5}(\omega,L_{\rm N},0)=\bar{m}_{6}(\omega,L_{\rm N},0)$ and thus
\begin{equation}
\begin{split}
f_{0,{\rm N}}^{r,{\rm E}}(x=L_{\rm N}/2,\omega)&=
-\frac{1}{2v_{F}}\frac{1}{{\rm sin}[\eta(\omega)-L_{\rm N}/\xi_{\omega}]}\,,\\
f_{3,{\rm N}}^{r,{\rm O}}(x=L_{\rm N}/2,\omega)&=0\,.
\end{split}
\end{equation}
Thus the OTE pairing is now completely suppressed for all energies within the energy gap. The ESE component, on the other hand, is still finite with multiple resonant peaks within the energy gap. These resonant peaks in the ESE magnitude directly correspond to the energies of the ABSs $\omega_{\pm,n}(\phi=0)=(v_{F}/2L)[2\pi(n+1/2)]$ bound within $\Delta$, as indicated by blue arrows in Fig.\,\ref{fig_SNS2}(c). 
Remarkably, the situation at $\phi=\pi$ implies $\bar{m}_{5}(\omega,L_{\rm N},\pi)=-\bar{m}_{5}(\omega,L_{\rm N},\pi)$, leading to instead
\begin{equation}
\begin{split}
f_{0,{\rm N}}^{r,{\rm E}}(x=L_{\rm N}/2,\omega)&=
0\,,\\
f_{3,{\rm N}}^{r,{\rm O}}(x=L_{\rm N}/2,\omega)&=\frac{1}{2iv_{F}}\frac{1}{{\rm cos}[\eta(\omega)-L_{\rm N}/\xi_{\omega}]}\,.
\end{split}
\end{equation}
Thus, at $\phi=\pi$ and in the middle of the N region, the OTE component is the only local pairing amplitude, since ESE is completely vanishing. Furthermore, the OTE resonant peaks reveal the emergence of the ABSs with energies $\omega_{\pm,n}(\phi=\pi)=(v_{F}/2L)[2\pi(n+1/2)\pm\pi]$, as indicated by red arrows in Fig.\,\ref{fig_SNS2}(d). 
Thus the peaks appearing in the subgap LDOS from the ABSs in the middle of the N region  can be directly associated with different pairing amplitudes. At $\phi=0$ the LDOS peaks are associated with pure ESE pairing, while at $\phi = \pi$ they are solely emerging due to the OTE pairing. Also, importantly, the zero energy bias peak at $\phi$ signaling the presence of Majorana fermions in the SNS junction is entirely associated with OTE pairing. This extends the results for a single pair of ABS in the short junction regime to include all subgap ABSs in longer junctions. 

\begin{figure}[htb]
\centering
\includegraphics[width=.49\textwidth]{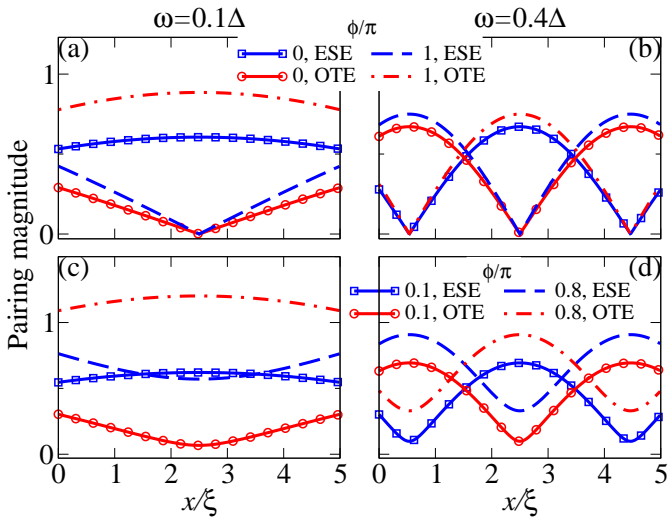} 
\caption{(Color online) ESE and OTE pairing magnitudes across the normal region of a SNS junction at energies $\omega=0.1\Delta$ (a,c)  and $\omega=0.4\Delta$  (b,d) for several different values of $\phi$. Parameters: $v_{F}=1$, $\Delta = 1$, $L_{\rm N}=5\xi$.}
\label{fig_SNS3}
\end{figure}

Above we found that for both short and long junctions it is only the OTE pairing that is intimately connected to the energy resonance at zero energy.
However, the dominance of the OTE pairing is not restricted to zero energy, $\phi=\pi$ phase, or the middle of the junction region. This can clearly be appreciated in Fig.\,\ref{fig_SNS3}, where we plot the spatial dependence of the ESE and OTE pairing magnitudes over the whole N region for energies away from zero and phase $\pi$. 
Although at the interface $x=0, L_N$, both ESE and OTE pairing magnitudes usually coexist at all phases, it is in the middle of the N region at $x=L_{\rm N}/2$, where OTE dominates over the completely suppressed ESE component for $\phi=\pi$.
The dominating OTE pairing extends over a notably finite region in the middle of the junction. It also survives to quite large subgap energies, $\omega \gtrsim 0.4\Delta$, and also for phase values.  Despite ESE often exhibiting a finite value in these regimes, the dominant behavior of OTE over ESE is clearly observed. Also notice how the spatial and energy dependences give rise to an oscillatory behavior of the pairing magnitudes given by Eqs.\,(\ref{SNS_N2}) as can be seen in Fig.\,\ref{fig_SNS3}. 
Overall this allows us to conclude that OTE pairing and its influence on the LDOS is not restricted neither to zero energy nor $\phi=\pi$.

\section{Conclusions}
\label{concl}
In this work we have analytically studied the emergence of odd-frequency superconductivity in NS, NSN, and SNS junctions at the edge of a 2DTI without magnetism or any other time-reversal symmetry breaking perturbations, where the S regions have spin-singlet $s$-wave superconducting pairing induced by proximity to an external conventional superconductor.
We have shown that odd-frequency mixed spin-triplet $s$-wave superconductivity does not require the presence of magnetic order but naturally arises at any NS interface as a result of breaking translation invariance in combination with the helicity of the 2DTI edge state. Moreover, we have clearly extended previous studies\cite{PhysRevB.75.134510,PhysRevB.76.054522,PhysRevLett.98.037003,Nagaosa12,PhysRevB.87.104513,PhysRevB.92.014508,PhysRevB.92.205424} and  established a one-to-one correspondence between the subgap LDOS and odd-frequency pairing in NS, NSN, and SNS junctions at the edge of 2DTI without any magnetism. These geometries are all suitable for LDOS as well as conductance measurements as has recently been demonstrated in experiments. \cite{PhysRevLett.109.186603,Yacoby14,vlad15,Bocquillon17}

For NS and NSN junctions, we have derived analytical expressions for both the pairing amplitudes and LDOS. Focusing on local, $s$-wave, pairing we have shown that both even-frequency, spin-singlet, even-parity (ESE) and odd-frequency, spin-triplet, even-parity (OTE) pairing is generally present in these systems. 
In the normal N region(s) the ESE and OTE magnitudes are equal and proportional to the Andreev reflection magnitude. 
In the superconducting S region the pairing amplitudes have two components arising from either the bulk or the interface(s). In the bulk of the S region, only the ESE pairing magnitude survives. However, there are both ESE and OTE interface components that develop an exponential decay into the bulk of the S region with the decay length set by the superconducting coherence length for small energies. 
Very close to the interfaces in the S region the OTE pairing becomes very dominant over an extremely suppressed ESE amplitude for energies well below the superconducting gap.
Moreover, the behavior of the low-energy LDOS in the S region close to the interface has the same dependence on the 
energy and distance from the interface as that of the OTE pairing. This allows us to associate the induced low-energy 
LDOS in the S region purely with OTE pairing.

In SNS junctions we have demonstrated an even stronger functional relationship between superconducting pairing and 
low-energy LDOS inside the N junction region. In this case the pairing amplitudes become dependent on the 
superconducting phase difference $\phi$ and on the length of the N region. 
In fact, we find that the condition giving the Andreev bound states (ABSs) energies in the junction is exactly the 
same as the condition generating resonant peaks in either the ESE and OTE pairing amplitudes, independent of junction length.
For short junctions, at $\phi=0$ the ESE pairing dominates over a completely suppressed OTE pairing, while at 
$\phi=\pi$ the ESE is instead zero and the OTE amplitude large. Notably, the pair of ABSs crosses zero at 
$\phi = \pi$, which is directly reflected in the OTE amplitude which has a resonant peak at zero energy. 
We have also shown that the supercurrent across such short junctions acquires its maximum value  
at $\phi=\pi$ as a result of the resonant peak at zero energy which is purely due to OTE pairing.
In long junctions more ABSs fit within the N region, but still all ABS energies correspond to resonant peaks in 
the pairing amplitudes. In this case we find the clearest distinction between ESE and OTE pairing in the middle of 
the N region. The ESE is completely suppressed at $\phi = \pi$, while the OTE is finite and has resonant peaks at 
each ABS energy. At $\phi=0$ the relation is reversed with ESE having resonances at the ABS energies. Dominant OTE 
pairing exists also at distances away from the middle of the junction and at phase differences away from $\phi=\pi$.

In summary, our findings show that the finite LDOS, as well as local conductance, at the interfaces of NS or NSN junctions at the edge of a 2DTI is entirely a consequence of pure OTE pairing. This results from the unique nature of  the helical edges of 2DTIs, where the Andreev reflection magnitude, strongly connected to LDOS and conductance, at the NS interface reaches its maximum for energies within the gap.\cite{PhysRevB.82.081303,PhysRevLett.109.186603} In SNS junctions the relationship is even stronger with the ABSs directly corresponding to peaks in the pairing amplitudes. In particular, the zero-energy ABS at $\phi = \pi$, protected by topology, is of complete OTE nature.

\section{Acknowledgements}
We thank P.~Burset, D.~Kuzmanovski and C.~Triola for motivating and helpful discussions.
This work was supported by the Swedish Research Council (Vetenskapsr\aa det), the G\"{o}ran Gustafsson Foundation, the Swedish Foundation for Strategic Research (SSF), and the Knut and Alice Wallenberg Foundation through the Wallenberg Academy Fellows program.

\bibliography{biblio}
\onecolumngrid
\clearpage
\twocolumngrid

\appendix

\section{Retarded and advanced Green's functions}
\label{AppG}
The structure of the Green's function $G^{r}$ is given by Eq.\,(\ref{Gr}) in the main text.
The elements $G_{ee,hh}^{r}$ we refer to as the regular parts of $G^{r}$, while $G_{eh,he}^{r}$ are the anomalous electron-hole parts. 
Electron-hole symmetry imposes for the BdG Hamiltonian $PH_{BdG}^{*}P^{\dagger}=-H_{BdG}$, while for the Green's function $P[G^{r}(x,x',\omega)]^{*}P^{\dagger}=-G^{r}(x,x',-\omega),$ where $P=\sigma_{y}\tau_{y}$.\cite{PhysRevB.92.100507} This therefore connects the two diagonal (off-diagonal) elements of $G^{r}$. 
The advanced Green's function, $G^{a}$, has the same matrix structure as the retarded function given by Eq.\,(\ref{Gr})
and can be calculated from incoming boundary conditions. Alternatively, we can use the relation between retarded and advanced Green's functions: $G^{a}(x,x',\omega)=[G^{r}(x',x,\omega)]^{\dagger}$.

We are interested in the pairing amplitudes, that is, the anomalous (electron-hole) part of the retarded Green's function, 
\begin{equation}
\label{AppG3}
G^{r}_{eh}(x,x',\omega)=
\begin{pmatrix}
[G^{r}_{eh}(x,x',\omega)]_{\uparrow\downarrow}&-[G^{r}_{eh}(x,x',\omega)]_{\uparrow\uparrow}\\
[G^{r}_{eh}(x,x',\omega)]_{\downarrow\downarrow}&-[G^{r}_{eh}(x,x',\omega)]_{\downarrow\uparrow}
\end{pmatrix}\,,
\end{equation}
where the minus signs arise due to the specific choice of basis.
In order to decompose the spin symmetry, we write the anomalous Green's function according to Eq.~\eqref{IsolateSpin} in the main text. Then we arrive at the pairing amplitudes $f^{r}_{i}$:
\begin{equation}
\label{eqf}
\begin{split}
f_{0}^{r}(x,x',\omega)&=\frac{[G^{r}_{eh}(x,x',\omega)]_{\uparrow\downarrow}-[G^{r}_{eh}(x,x',\omega)]_{\downarrow\uparrow}}{2}\,,\\
f_{3}^{r}(x,x',\omega)&=\frac{[G^{r}_{eh}(x,x',\omega)]_{\uparrow\downarrow}+[G^{r}_{eh}(x,x',\omega)]_{\downarrow\uparrow}}{2}\,,\\
f_{1}^{r}(x,x',\omega)&=\frac{-[G^{r}_{eh}(x,x',\omega)]_{\uparrow\uparrow}+[G^{r}_{eh}(x,x',\omega)]_{\downarrow\downarrow}}{2}\,,\\
f_{2}^{r}(x,x',\omega)&=i\frac{[G^{r}_{eh}(x,x',\omega)]_{\uparrow\uparrow}+[G^{r}_{eh}(x,x',\omega)]_{\downarrow\downarrow}}{2}\,.
\end{split}
\end{equation}
From these equations we observe that $f_{0}^{r}$ acquires a minus sign under the exchange of spins and is thus an odd function under spin exchange, while $f^{r}_{1,2,3}$ acquire a plus sign and are even under spin exchange. 
 $f^{r}_{0}$ is therefore referred to as the spin-singlet component, while $f^{R}_{1,2,3}$ are the spin-triplet components.
 
\subsection{Antisymmetry}
\label{antify}
The anomalous Green's function represents the wave function of a two-electron system which must obey antisymmetry upon the simultaneous exchange of spins ($\uparrow\leftrightarrow\downarrow$), spatial coordinates ($x\leftrightarrow x'$) and time (or energy/frequency) coordinates ($t\leftrightarrow t'$ or $\omega\rightarrow-\omega$). 
In this work, we use retarded and advanced Green's functions which do  not respect symmetry under frequency. These Green's functions are only partially defined on the time axis and therefore, when the sign of the frequency changes (or time), we should pass from one to the other,\cite{PhysRevB.92.100507,PhysRevB.92.205424}
\begin{equation}
\label{AntiGg}
G^{r}_{s_{1}s_{2}}(x,x',\omega)=-G^{a}_{s_{2}s_{1}}(x',x,-\omega)\,.
\end{equation}
By using Eqs.\,(\ref{AppG3}) and (\ref{AntiGg}) we get
\begin{equation}
\begin{split}
[G^{r}_{eh}(x,x',\omega)]_{\uparrow\downarrow}&=[G^{a}_{eh}(x',x,-\omega)]_{\downarrow\uparrow}\,,\\
[G^{r}_{eh}(x,x',\omega)]_{\uparrow\uparrow}&=-[G^{a}_{eh}(x',x,-\omega)]_{\uparrow\uparrow}\,,\\
[G^{r}_{eh}(x,x',\omega)]_{\downarrow\downarrow}&=-[G^{a}_{eh}(x',x,-\omega)]_{\downarrow\downarrow}\,,\\
[G^{r}_{eh}(x,x',\omega)]_{\downarrow\uparrow}&=[G^{a}_{eh}(x',x,-\omega)]_{\uparrow\downarrow}\,.
\end{split}
\end{equation}
Similarly, we arrive at the conditions of antisymmetry for the spin-singlet and triplet amplitudes
\begin{equation}
\label{antif}
\begin{split}
f^{r}_{0}(x,x',\omega)&=f_{0}^{a}(x',x,-\omega)\,,\\
f^{r}_{1}(x,x',\omega)&=-f_{1}^{a}(x',x,-\omega)\,,\\
f^{r}_{2}(x,x',\omega)&=-f_{2}^{a}(x',x,-\omega)\,,\\
f^{r}_{3}(x,x',\omega)&=-f_{3}^{a}(x',x,-\omega)\,.
\end{split}
\end{equation}
These equations represent the full antisymmetry conditions under the exchange of spin, spatial coordinates, and frequency imposed by Fermi-Dirac statistics on the pairing amplitudes and they are presented in the main text as Eqs.\,(\ref{antif_main}).

\section{NS junction}
\label{NSApp}
We model a NS junction at the metallic edge of a 2DTI by considering a step-like profile of the induced pairing potential, with the normal region at $x<0$ and the superconducting region at  $x>0$, 
\begin{equation}
\Delta(x)=\theta(x)\Delta=\begin{cases}
   0 \,,  & x<0, \\
    \Delta \,, & x>0\,,
\end{cases}
\end{equation}
where we can set the overall superconducting phase to zero. 

\subsection{Scattering processes}
In general, there are four different scattering processes at the NS interface which read
\begin{equation}
\label{normalw}
\begin{split}
\Psi_{1}(x)&=
\begin{cases}
     \phi_{1}^{N}\,{\rm e}^{ik_{e}x}+a_{1}\phi_{3}^{N}\,{\rm e}^{ik_{h}x} +b_{1}\phi_{2}^{N}\,{\rm e}^{-ik_{e}x},\,x<0&  \\
    c_{1}  \phi_{1}^{S}\,{\rm e}^{ik_{e}^{S}x} +d_{1}  \phi_{4}^{S}\,{\rm e}^{-ik_{h}^{S}x},\,x>0 & 
\end{cases}\\
\Psi_{2}(x)&=
\begin{cases}
     \phi_{4}^{N}\,{\rm e}^{-ik_{h}x}+a_{2}\phi_{2}^{N}\,{\rm e}^{-ik_{e}x} +b_{2}\phi_{3}^{N}\,{\rm e}^{ik_{h}x},\,x<0& \\
    c_{2}  \phi_{4}^{S}\,{\rm e}^{-ik_{h}^{S}x} +    d_{2}  \phi_{1}^{S}\,{\rm e}^{ik_{e}^{S}x}, \,x>0 & 
\end{cases}\\
\Psi_{3}(x)&=
\begin{cases}
    c_{3}  \phi_{2}^{N}\,{\rm e}^{-ik_{e}x} +d_{3}  \phi_{3}^{N}\,{\rm e}^{ik_{h}x},\, x<0&\\
     \phi_{2}^{S}\,{\rm e}^{-ik_{e}^{S}x}+a_{3}\phi_{4}^{S}\,{\rm e}^{-ik_{h}^{S}x}+b_{3}\phi_{1}^{S}\,{\rm e}^{ik_{e}^{S}x},\,x>0&  
\end{cases}\\
\Psi_{4}(x)&=
\begin{cases}
    c_{4}  \phi_{3}^{N}\,{\rm e}^{ik_{h}x} +    d_{4}  \phi_{2}^{N}\,{\rm e}^{-ik_{e}x},\,x<0  & \\
\phi_{3}^{S}\,{\rm e}^{ik_{h}^{S}x}+a_{4}\phi_{1}^{S}\,{\rm e}^{ik_{e}^{S}x}+b_{4}\phi_{4}^{S}\,{\rm e}^{-ik_{h}^{S}x},\,x>0&
\end{cases}
\end{split}
\end{equation}
while the conjugated processes are
\begin{equation}
\label{conjuw}
\begin{split}
\tilde{\Psi}_{1}(x')&=
\begin{cases}
    \tilde{ \phi}_{1}^{N}\,{\rm e}^{ik_{e}x'}+\tilde{a}_{1}\tilde{\phi}_{3}^{N}{\rm e}^{ik_{h}x'} +\tilde{b}_{1}\tilde{\phi}_{2}^{N}{\rm e}^{-ik_{e}x'},\, x<0& \\
    \tilde{c}_{1} \tilde{ \phi}_{1}^{S}{\rm e}^{ik_{e}^{S}x'} +  \tilde{d}_{1} \tilde{ \phi}_{4}^{S}\,{\rm e}^{-ik_{h}^{S}x'},\,x>0  &
\end{cases}\\
\tilde{\Psi}_{2}(x')&=
\begin{cases}
   \tilde{\phi}_{4}^{N}{\rm e}^{-ik_{h}x'}+\tilde{a}_{2}\tilde{\phi}_{2}^{N}{\rm e}^{-ik_{e}x'} +\tilde{b}_{2}\tilde{\phi}_{3}^{N}{\rm e}^{ik_{h}x'},\,x<0&  \\
 \tilde{c}_{2} \tilde{\phi}_{4}^{S}{\rm e}^{-ik_{h}^{S}x'} + \tilde{d}_{2} \tilde{ \phi}_{1}^{S}{\rm e}^{ik_{e}^{S}x'},\,x>0 &
\end{cases}\\
\tilde{\Psi}_{3}(x')&=
\begin{cases}
    \tilde{c}_{3}  \tilde{\phi}_{2}^{N}{\rm e}^{-ik_{e}x'} +\tilde{d}_{3} \tilde{ \phi}_{3}^{N}{\rm e}^{ik_{h}x'}, \,x<0 &\\
     \tilde{\phi}_{2}^{S}{\rm e}^{-ik_{e}^{S}x'}+\tilde{a}_{3}\tilde{\phi}_{4}^{S}{\rm e}^{-ik_{h}^{S}x'}+\tilde{b}_{3}\tilde{\phi}_{1}^{S}{\rm e}^{ik_{e}^{S}x'},\,x>0& 
\end{cases}\\
\tilde{\Psi}_{4}(x')&=
\begin{cases}
    \tilde{c}_{4} \tilde{ \phi}_{3}^{N}{\rm e}^{ik_{h}x'} +    \tilde{d}_{4}  \tilde{\phi}_{2}^{N}{\rm e}^{-ik_{e}x'},\, x<0  &\\
    \tilde{ \phi}_{3}^{S}{\rm e}^{ik_{h}^{S}x'}+\tilde{a}_{4}\tilde{\phi}_{1}^{S}\,{\rm e}^{ik_{e}^{S}x'}+\tilde{b}_{4}\tilde{\phi}_{4}^{S}{\rm e}^{-ik_{h}^{S}x'},\, x>0\,,& 
\end{cases}
\end{split}
\end{equation}
where 
\begin{equation}
\label{phiNS}
\begin{split}
\phi_{1}^{N}&=\begin{pmatrix}
1,
0,
0,
0
 \end{pmatrix}^{T},\\
\phi_{2}^{N}&=\begin{pmatrix}
0,
1,
0,
0
 \end{pmatrix}^{T},\\
\phi_{3}^{N}&=\begin{pmatrix}
0,
0,
1,
0
 \end{pmatrix}^{T},\\
\phi_{4}^{N}&=\begin{pmatrix}
0,
0,
0,
1
 \end{pmatrix}^{T},
 \end{split}\,\quad
\begin{split}
 \phi_{1}^{S}&=\begin{pmatrix}
u,
0,
v,
0
 \end{pmatrix}^{T},\\
\phi_{2}^{S}&=\begin{pmatrix}
0,
u,
0,
v
 \end{pmatrix}^{T},\\
\phi_{3}^{S}&=\begin{pmatrix}
v,
0,
u,
0
 \end{pmatrix}^{T},\\
\phi_{4}^{S}&=\begin{pmatrix}
0,
v,
0,
u
 \end{pmatrix}^{T},\,
\end{split}
\end{equation}
are wave functions of $H_{BdG}(k)$, while 
\begin{equation}
\label{tildephiNS}
\begin{split}
\tilde{\phi}_{1}^{N}&=\begin{pmatrix}
0,
1,
0,
0
\end{pmatrix}^{T},\\
 \tilde{\phi}_{2}^{N}&=\begin{pmatrix}
1,
0,
0,
0
\end{pmatrix}^{T},\\
\tilde{\phi}_{3}^{N}&=\begin{pmatrix}
0,
0,
0,
1
\end{pmatrix}^{T},\\
 \tilde{\phi}_{4}^{N}&=\begin{pmatrix}
0,
0,
1,
0
\end{pmatrix}^{T},
\end{split}\,\quad
\begin{split}
\tilde{\phi}_{1}^{S}&=
\begin{pmatrix}
0,
u,
0,
v
\end{pmatrix}^{T},\\
\tilde{\phi}_{2}^{S}&=
\begin{pmatrix}
u,
0,
v,
0
\end{pmatrix}^{T},\\
\tilde{\phi}_{3}^{S}&=
\begin{pmatrix}
0,
v,
0,
u
\end{pmatrix}^{T},\\
\tilde{\phi}_{4}^{S}&=
\begin{pmatrix}
v,
0,
u,
0
\end{pmatrix}^{T},\,
\end{split}
\end{equation}
are the wave functions of the conjugated Hamiltonian $\tilde{H}_{BdG}(k)$.  
Notice that the conjugated scattering processes are constructed after solving the eigenvalue problem for $\tilde{H}_{BdG}(k)=H_{BdG}^{*}(-k)=H_{BdG}^{T}(-k)$ instead of Eq.\,\eqref{H2DSC} in the main text.
In these equations we have used the following relations:
\begin{equation}
\begin{split}
k_{e,h}&=(\mu\pm \omega)/v_{F}\,,\\
k_{e,h}^{S}&=(\mu\pm\sqrt{\omega^{2}-\Delta^{2}})/v_{F}=k_{\mu}\pm k(\omega)\,,\\
 u,v&=\sqrt{\frac{1}{2}\bigg[1\pm\frac{\sqrt{\omega^{2}-\Delta^{2}}}{\omega} \bigg]}\,.
\end{split}
 \end{equation}
In this work we mainly focus on energies within $\Delta$ and then 
 $k_{e,h}^{S}=k_{\mu}\pm i\kappa(\omega)$, with $\kappa(\omega)=\sqrt{\Delta^{2}-\omega^{2}}/v_F$ and the coherence factors can be written as $u=\sqrt{\Delta/2\omega}\,{\rm e}^{i\eta(\omega)/2}$ and $v=\sqrt{\Delta/2\omega}\,{\rm e}^{-i\eta(\omega)/2}$, where $\eta(\omega)={\rm arccos}(\omega/\Delta)$.
 
At the NS interface, $\Psi_{1}$ represents the following process: 
an incoming electron (right-moving with spin up) from the N region with wave function $\phi_{1}^{N}\,{\rm e}^{ik_{e}x}$ 
experiences reflection and transmission at the NS interface with certain probabilities.
It can be reflected into a left-moving electron with spin-down with wave function 
$\phi_{2}^{N}\,{\rm e}^{-ik_{e}x}$ and amplitude $b_{1}$ or Andreev reflected into a left-moving hole with spin down with wave function $\phi_{3}^{N}\,{\rm e}^{ik_{h}x}$ and amplitude $a_{1}$, or transmitted to into the S region in the form of a right-moving quasielectron with wave function 
$\phi_{1}^{S}\,{\rm e}^{ik_{e}^{S}x}$ and amplitude $c_{1}$ or a right-moving quasihole with wave function $\phi_{4}^{S}\,{\rm e}^{-ik_{h}^{S}x}$ and amplitude $d_{1}$. 
Thus, $a_{1}$, $b_{1}$, $c_{1}$ and $d_{1}$ are the amplitudes of reflection into a hole (Andreev reflection), reflection into an electron (normal reflection), transmission into an electron and transmission into a hole. 
Likewise, $\Psi_{2,3,4}$ correspond to scattering processes for an incoming hole from the N region and incoming electron or incoming hole from the S region, respectively. 
 
The processes $\Psi_{i}$ and $\tilde{\Psi}_{i}$ are fully determined after finding the coefficients $a_{i}$, $b_{i}$, $c_{i}$ and $d_{i}$. These in turn are calculated by matching the functions $\Psi_{i}$ at the NS interface $x=0$,
\begin{equation}
\label{NSsharp1}
\begin{split}
\Big[\Psi_{i}(x>0)\Big]_{x=0}&=\Big[\Psi_{i}(x<0)\Big]_{x=0}\,.
\end{split}
\end{equation}
Each $\Psi_{i}$ is a four column vector and therefore provides four equations. At the end we have a system of $16$ equations for the $16$ unknown coefficients $a_{i}$, $b_{i}$, $c_{i}$ and $d_{i}$, which is generally solvable and completely determines the  scattering states $\Psi_{i}$.  
The same holds for the conjugated processes $\tilde{\Psi}_{i}$. 
Notice that we here have written the scattering wave functions, Eqs.\,(\ref{normalw}-\ref{conjuw}), 
in a general form. However, the processes with amplitudes $b_{i}$ and $d_{i}$ are forbidden by helicity 
of the 2DTI edge; an incident electron can be only reflected as a hole by a superconducting barrier or 
transmitted as an electron through it. 
Thus, normal reflection and non-local Andreev transmission are forbidden by helicity conservation\cite{PhysRevB.82.081303,PhysRevB.88.075401} and $b_{i}=d_{i}=0$.  For the Andreev reflection amplitudes we generally also obtain $a_{1,2}=v/u=-a_{3,4}$.

\subsection{Green's functions and pairing amplitudes}
\label{NSApp1}

\subsubsection{Normal region}
\label{NSApp2}
After finding $\Psi_{i}$ and $\tilde{\Psi}_{i}$, given by Eqs.\,(\ref{normalw}) and (\ref{conjuw}), we construct the retarded Green's functions using Eq.\,(\ref{GFUNCTION}) in the main text.
In the normal (N) region, we obtain the following expressions for the regular and anomalous parts
\begin{equation}
\label{GFNS_N}
\begin{split}
G_{ee}^{r}(x,x',\omega)
&=
\frac{1}{iv_{f}}
\begin{pmatrix}
G_{ee,\uparrow\uparrow}^{r}&0\\
0&G_{ee,\downarrow\downarrow}^{r}
\end{pmatrix},\\
G_{hh}^{r}(x,x',\omega)
&=\frac{1}{iv_{f}}\begin{pmatrix}
G_{hh,\downarrow\downarrow}^{r}&0\\
0&G_{hh,\uparrow\uparrow}^{r}\\
\end{pmatrix},\\
G_{eh}^{r}(x,x',\omega)
&=
\frac{1}{iv_{f}}\begin{pmatrix}
0&0\\
0&a_{1}(\omega){\rm e}^{-i(k_{e}x-k_{h}x')}\\
\end{pmatrix},\\
G_{he}^{r}(x,x',\omega)
&=
\frac{1}{iv_{f}}\begin{pmatrix}
a_{1}(\omega){\rm e}^{i(k_{h}x-k_{e}x')}&0\\
0&0
\end{pmatrix},
\end{split}
\end{equation}
with $G_{ee,\uparrow\uparrow}^{r}=\theta(x-x'){\rm e}^{ik_{e}(x-x')}$, 
$G_{ee,\downarrow\downarrow}^{r}=\theta(x'-x){\rm e}^{-ik_{e}(x-x')}$, 
$G_{hh,\downarrow\downarrow}^{r}=\theta(x'-x){\rm e}^{ik_{h}(x-x')}$,
$G_{hh,\uparrow\uparrow}^{r}=\theta(x-x'){\rm e}^{-ik_{h}(x-x')}$ and 
$a_{1}=v/u$ the Andreev reflection amplitude for an incident electron from N. For energies within the gap $a_{1}(\omega)={\rm e}^{-i\eta(\omega)}$.
Now, by using $G^{r}_{ee}$, we calculate the LDOS in the N region following Eq.\,(\ref{LDOS}) in the main text and get
\begin{equation}
\label{DOS_NS_N}
\begin{split}
\rho_{{\rm N}}(x,\omega)
&=\frac{1}{\pi v_{F}}\,,
\end{split}
\end{equation}
where we have used ${\rm lim}_{x\rightarrow x'}\theta(x-x')=1/2$. 
Notice how the LDOS is independent of both energy and position, as is expected for the helical edge state in a 2DTI.

By decomposing the spin symmetry of the anomalous part of the retarded and advanced Green's functions, employing Eq.\,(\ref{IsolateSpin}), we get the pairing amplitudes
\begin{equation}
\label{f_NS_N}
\begin{split}
f_{0,{\rm N}}^{r}(x,x',\omega)&=\frac{a_{1}(\omega)}{2i v_{F}}
{\rm e}^{-i\frac{\mu(x-x')+\omega(x+x')}{v_{F}}}
,\\
f_{3,{\rm N}}^{r}(x,x',\omega)&=-f_{0}^{r}(x,x',\omega),
\end{split}
\end{equation}
and $f_{1,2,{\rm N}}^{r}=0$.
As discussed in Sec.\,\ref{sect1}, the pairing amplitudes must obey Fermi-Dirac statistics. We use Eqs.\,(\ref{antif}) and check the antisymmetry of previous pairing functions.
For example, for the singlet component we have
\begin{equation}
\begin{split}
f^{a}_{0,{\rm N}}(x',x,-\omega)&=-\frac{a_{1}^{*}(-\omega)}{2iv_{F}}
{\rm e}^{i\frac{\mu(x'-x)-\omega(x+x')}{v_{F}}}
\,,\\
&=\frac{a_{1}(\omega)}{2iv_{F}}
{\rm e}^{-i\frac{\mu(x-x')+\omega(x+x')}{v_{F}}}
\,,\\
&=f_{0}^{r}(x,x',\omega)\,,
\end{split}
\end{equation}
where we have used $a_{1}^{*}(-\omega)={\rm e}^{i\eta(-\omega)}=-{\rm e}^{-i\eta(\omega)}=-a_{1}(\omega)$. Thus, we conclude that $f_{0}^{r}$ is antisymmetric. 
Likewise, the triplet component is fully antisymmetric and obeys the relation given by Eq.\,(\ref{antif}). Notice that in order to check antisymmetry we have to use the advanced pairing functions as discussed in Appendix \ref{AppG}. 
We now use Eqs.\,(\ref{Odd_Even_text}) in order to decompose into the even and odd-frequency components
\begin{equation}
\label{f_NS_N2}
\begin{split}
f^{r,{\rm E}}_{0,{\rm N}}(x,x',\omega)&=\frac{a_{2}(\omega)}{2iv_{F}}\,C_{xx'}\,{\rm e}^{-i\frac{\omega(x+x')}{v_{F}}},\\
f^{r,{\rm O}}_{0,{\rm N}}(x,x',\omega)&=-\frac{a_{2}(\omega)}{2v_{F}}\,S_{xx'}\,{\rm e}^{-i\frac{\omega(x+x')}{v_{F}}},\\
f^{r,{\rm E}}_{3,{\rm N}}(x,x',\omega)&=\frac{a_{2}(\omega)}{2v_{F}}\,S_{xx'}\,{\rm e}^{-i\frac{\omega(x+x')}{v_{F}}},\\
f^{r,{\rm O}}_{3,{\rm N}}(x,x',\omega)&=-\frac{a_{2}(\omega)}{2iv_{F}}\,C_{xx'}\,{\rm e}^{-i\frac{\omega(x+x')}{v_{F}}},
\end{split}
\end{equation}
where $C_{xx'}={\rm cos}[k_{\mu}(x-x')]$, $S_{xx'}={\rm sin}[k_{\mu}(x-x')]$, and $a_{2}=a_{1}$ the Andreev reflection amplitude. These are Eqs.~\eqref{f_NS_N2_Main} given in the main text.

Before going further it is worth to point out the following. To check the antisymmetry of the retarded even- and odd-frequency pairing amplitudes is not trivial and it is important to write down their respective advances functions. We have explicitly verified that the given expressions  follow the antisymmetry relations. For a quick check, however, we can focus on the parity in space which follows directly from ${\rm cos}[k_{\mu}(x-x')]$ and ${\rm sin}[k_{\mu}(x-x')]$. Since the spin is already explicit this means the only symmetry left is that of the frequency. For example, for OSO we see that ${\rm sin}[k_{\mu}(x-x')]$ makes the amplitude odd in space and since it is a spin singlet, the only possibility is for an odd-frequency dependence.

\subsubsection{Superconducting region}
\label{NSApp2}
In the superconducting region (S) we proceed similarly as in N. We obtain the retarded Green's function, which for energies below the gap $\Delta$, reads
\begin{widetext}
\begin{equation}
\label{GFNS_S}
\begin{split}
G^{r}_{ee,\uparrow\uparrow}(x,x',\omega)&=Z(\omega)
\,{\rm e}^{ik_{\mu}(x-x')}
\Big[
\tilde{a}_{3}(\omega)\,{\rm e}^{-\kappa(\omega)(x+x')}
+\,{\rm e}^{-\kappa(\omega)|x-x'|}N(x,x',\omega)\Big]\,,\\
G^{r}_{ee,\downarrow\downarrow}(x,x',\omega)&=
Z(\omega)
\,{\rm e}^{-ik_{\mu}(x-x')}
\Big[\tilde{a}_{3}(\omega)\,{\rm e}^{-\kappa(\omega)(x+x')}
+{\rm e}^{-\kappa(\omega)|x-x'|}N(x',x,\omega)\Big]
\,,\\
G^{r}_{hh,\downarrow\downarrow}(x,x',\omega)&=
Z(\omega)
\,{\rm e}^{ik_{\mu}(x-x')}
\Big[\tilde{a}_{3}(\omega)\,{\rm e}^{-\kappa(\omega)(x+x')}
+{\rm e}^{-\kappa(\omega)|x-x'|}N(x',x,\omega)\Big]
\,,\\
G^{r}_{hh,\uparrow\uparrow}(x,x',\omega)&=Z(\omega)
\,{\rm e}^{-ik_{\mu}(x-x')}
\Big[
\tilde{a}_{3}(\omega)\,{\rm e}^{-\kappa(\omega)(x+x')}
+\,{\rm e}^{-\kappa(\omega)|x-x'|}N(x,x',\omega)\Big]\,,\\
G^{r}_{eh,\uparrow\downarrow}(x,x',\omega)&=
Z(\omega)
\,{\rm e}^{ik_{\mu}(x-x')}
\Big[
\frac{u}{v}\tilde{a}_{3}(\omega)\,{\rm e}^{-\kappa(\omega)(x+x')}
+\,{\rm e}^{-\kappa(\omega)|x-x'|}\Big]\,,\\
G^{r}_{eh,\downarrow\uparrow}(x,x',\omega)&=
Z(\omega)
\,{\rm e}^{-ik_{\mu}(x-x')}
\Big[
\frac{v}{u}\tilde{a}_{3}(\omega)\,{\rm e}^{-\kappa(\omega)(x+x')}
+\,{\rm e}^{-\kappa(\omega)|x-x'|}\Big]\,,\\
G^{r}_{he,\downarrow\uparrow}(x,x',\omega)&=
Z(\omega)
\,{\rm e}^{ik_{\mu}(x-x')}
\Big[
\frac{v}{u}\tilde{a}_{3}(\omega)\,{\rm e}^{-\kappa(\omega)(x+x')}
+\,{\rm e}^{-\kappa(\omega)|x-x'|}\Big]\,,\\
G^{r}_{he,\uparrow\downarrow}(x,x',\omega)&=
Z(\omega)
\,{\rm e}^{-ik_{\mu}(x-x')}
\Big[
\frac{u}{v}\tilde{a}_{3}(\omega)\,{\rm e}^{-\kappa(\omega)(x+x')}
+\,{\rm e}^{-\kappa(\omega)|x-x'|}\Big]\,,\\
G^{r}_{eh(he),\uparrow\uparrow,\downarrow\downarrow}(x,x',\omega)&=0\,,\quad
G^{r}_{ee(hh),\uparrow\downarrow,\downarrow\uparrow}(x,x',\omega)=0\,,\\
\end{split}
\end{equation}
\end{widetext}
where $N(x,x',\omega)=\theta(x-x')(u/v)+\theta(x'-x)(v/u)$, $Z(\omega)=(1/iv_{F})/[{\rm e}^{i\eta(\omega)}-{\rm e}^{-i\eta(\omega)}]$, and $\tilde{a}_{3}=a_{3}=-a_{2}$.
Notice that, while in the anomalous parts (eh, he) only mixed spin components are finite, in the regular parts (ee, hh) we also obtain finite equal spin components. If we were to consider finite magnetic order in the system the situation would dramatically change with additional off-diagonal (mixed spin) terms in the regular part and equal spin terms in the anomalous components.\cite{PhysRevB.92.100507}
By using the regular electron-electron part of $G^{r}$, given by the first two expressions of Eqs.\,(\ref{GFNS_S}), we obtain the LDOS in the superconducting region,
\begin{equation}
\label{DOS_NS_S}
\begin{split}
\rho_{\rm S}(x,\omega)
&={\rm Im}\Big\{\frac{i}{\pi v_{F}}\Big[ \bar{\rho}+(1- \bar{\rho}){\rm e}^{ik(\omega)2x}\Big]\Big\}\,,
\end{split}
\end{equation}
where $\bar{\rho}(\omega)=\omega/\sqrt{\omega^{2}-\Delta^{2}}$, and $k(\omega)=i\kappa(\omega)$.
Within the gap, $\omega^{2}<\Delta^{2}$, this reduces to
\begin{equation}
\label{DOS_NS_S2}
\begin{split}
\rho_{\rm S}(x,|\omega|<|\Delta|)
&=\frac{1}{\pi v_{F}}{\rm e}^{-\kappa(\omega)2x}\,,
\end{split}
\end{equation}
which is given as Eq.\,(\ref{DOS_NS_S_main}) in the main text.

The pairing amplitudes are calculated by decomposing the spin components according to Eq.\,(\ref{IsolateSpin}) and we arrive at
\begin{widetext}
\begin{equation}
\label{pairingNS_S}
\begin{split}
f_{0,S}^{r}(x,x',\omega)&=
Z(\omega){\rm e}^{-\kappa(\omega)|x-x'|} C_{xx'}
+
Z(\omega)\frac{\tilde{a}_{3}(\omega)}{2}{\rm e}^{-\kappa(\omega)(x+x')}
\Big[
{\rm e}^{i\eta(\omega)}{\rm e}^{ik_{\mu}(x-x')}+{\rm e}^{-i\eta(\omega)}{\rm e}^{-ik_{\mu}(x-x')}
\Big]\,,\\
f_{3,S}^{r}(x,x',\omega)&=
Z(\omega){\rm e}^{-\kappa(\omega)|x-x'|} iS_{xx'}
+
Z(\omega)\frac{\tilde{a}_{3}(\omega)}{2}
{\rm e}^{-\kappa(\omega)(x+x')}
\Big[
{\rm e}^{i\eta(\omega)}{\rm e}^{ik_{\mu}(x-x')}-{\rm e}^{-i\eta(\omega)}{\rm e}^{-ik_{\mu}(x-x')}
\Big]\,,\\
\end{split}
\end{equation}
\end{widetext}
with $f_{1,2,S}^{r}=0$.
It is important to notice that the pairing amplitudes given by Eqs.\,(\ref{pairingNS_S}) contain two terms, which arise from different parts in the S region. The first term is proportional to ${\rm e}^{-\kappa(\omega)|x-x'|}$ and is associated with the bulk deep inside the S region. Indeed, locally $(x = x')$, such a term becomes independent of the space coordinate. The second term is proportional to the Andreev reflection amplitude $a_{3}$ and therefore arises due the presence of the NS interface. This term also keeps a spatial dependence even locally that gives the decay into the S region.
Thus, in the S region, the pairing amplitudes can be written with
as bulk (B)  and interface (I) components
\begin{equation}
\label{pairingNS_S2}
\begin{split}
f_{0,{\rm B}}^{r}(x,x',\omega)&=
Z(\omega){\rm e}^{-\kappa(\omega)|x-x'|} C_{xx'}\,,\\
f_{3,{\rm B}}^{r}(x,x',\omega)&=
Z(\omega){\rm e}^{-\kappa(\omega)|x-x'|} iS_{xx'}\,,\\
f_{0,{\rm I}}^{r}(x,x',\omega)&=
Z(\omega)\frac{\tilde{a}_{3}(\omega)}{2}{\rm e}^{-\kappa(\omega)(x+x')}
K_{+}\,,\\
f_{3,{\rm I}}^{r}(x,x',\omega)&=
Z(\omega)\frac{\tilde{a}_{3}(\omega)}{2}{\rm e}^{-\kappa(\omega)(x+x')}
K_{-}\,,\\
\end{split}
\end{equation}
where we have used $K_{\pm}(\omega,x,x')=
{\rm e}^{i\eta(\omega)}{\rm e}^{ik_{\mu}(x-x')}\pm{\rm e}^{-i\eta(\omega)}{\rm e}^{-ik_{\mu}(x-x')}$. 
We have explicitly checked that all these pairing amplitudes are fully antisymmetric, obeying Eqs.\,(\ref{antif}). For example, for $f_{3,{\rm I}}^{r}$ we find
\begin{equation}
\label{TSC_NS}
\begin{split}
f_{3,{\rm I}}^{a}(x',x,-\omega)&=
{Z}^\ast(-\omega)\frac{\tilde{a}^{*}_{3}(-\omega)}{2}{\rm e}^{-\kappa(-\omega)(x+x')}\\
&\times K_{-}(-\omega,x',x)\,,\\
&=-Z(\omega)\frac{\tilde{a}_{3}(\omega)}{2}{\rm e}^{-\kappa(\omega)(x+x')}\\
&\times K_{-}(\omega,x,x')\,,\\
&=-f_{3,{\rm I}}^{r}(x,x',\omega)
\end{split}
\end{equation}
where we have used $\tilde{a}_{3}^{*}(-\omega)=-a_{3}(\omega)$, ${Z}^\ast(-\omega)=Z(\omega)$ and $K_{-}(-\omega,x',x)=K_{-}(\omega,x,x')$. We have also used the advanced pairing function $f^{a}_{3,{\rm I}}$ which is  calculated from the relation between retarded and advanced Green's functions $G^{a}(x,x',\omega)=[G^{r}(x',x,\omega)]^{\dagger}$.

We now use Eqs.\,(\ref{Odd_Even_text}) in the main text to isolate the even- and odd-frequency components, which gives in the bulk of the S region
\begin{equation}
\label{f_NS_Sx}
\begin{split}
f^{r,{\rm O}}_{0,3,{\rm B}}(x,x',\omega)&=0\,,\\
f^{r,{\rm E}}_{0,{\rm B}}(x,x',\omega)&=Z(\omega){\rm e}^{-\kappa(\omega)|x-x'|}\,C_{xx'}\,,\\
f^{r,{\rm E}}_{3,{\rm B}}(x,x',\omega)&=iZ(\omega){\rm e}^{-\kappa(\omega)|x-x'|}\,S_{xx'}\,.\\
\end{split}
\end{equation}
and close to the NS interface on the S side
\begin{equation}
\label{f_NS_Sy}
\begin{split}
f^{r,{\rm E}}_{0,{\rm I}}(x,x',\omega)&=\frac{a_{3}(\omega)}{2iv_{F}}{\rm e}^{-\kappa(\omega)(x+x')}
B(\omega)
C_{xx'}\,,\\
f^{r,{\rm O}}_{0,{\rm I}}(x,x',\omega)&=\frac{a_{3}(\omega)}{2v_{F}}{\rm e}^{-\kappa(\omega)(x+x')}\,S_{xx'}\,,\\
f^{r,{\rm E}}_{3,{\rm I}}(x,x',\omega)&=\frac{a_{3}(\omega)}{2v_{F}}{\rm e}^{-\kappa(\omega)(x+x')}
B(\omega)
S_{xx'}\,,\\
f^{r,{\rm O}}_{3,{\rm I}}(x,x',\omega)&=\frac{a_{3}(\omega)}{2iv_{F}}{\rm e}^{-\kappa(\omega)(x+x')}\,C_{xx'}\,,\\
\end{split}
\end{equation}
where $B(\omega)=[{\rm e}^{i\eta(\omega)}+{\rm e}^{-i\eta(\omega)}]/[{\rm e}^{i\eta(\omega)}-{\rm e}^{-i\eta(\omega)}]\,$. 
Notice that in the bulk we only obtain trivial even-frequency spin-singlet and -triplet components, namely the ESE and ETO pairing amplitudes.
At the interface, however, hosts all symmetry classes (ESE, OSO, ETO, and OTE). Previous two sets of equations, for the bulk and interface, correspond to Eqs.~(\ref{f_NS_Sx_main}-\ref{f_NS_Sy_main}) in the main text.

\section{NSN}
\label{NSNApp}
Here we consider a NSN junction, where the S region has a finite length $L_{\rm S}$ restricted to $0<x<L_{\rm S}$, i.e.,~
\begin{equation}
\label{DeltaNSN}
\Delta(x)=\begin{cases}
   0 \,,  & x<0\,, \\
    \Delta \,, & 0<x<L_{\rm S}\,,\\
     0 \,,  & x>0\,.
\end{cases}
\end{equation}

\subsection{Scattering processes}
The four scattering processes in a NSN junction read
\begin{equation}
\label{eqnsn1}
\begin{split}
\Psi_{1}(x)&=
\begin{cases}
      \phi_{1}^{N}{\rm e}^{ik_{e}x}+a_{1}\phi_{3}^{N}{\rm e}^{ik_{h}x}+b_{1}\phi_{2}^{N}{\rm e}^{-ik_{e}x},\, x<0& \\
       \sum_{i}p_{i}\phi_{i}^{S}{\rm e}^{ik_{i}^{S}x},\,0<x<L_{\rm S}& \\
   c_{1}\phi_{1}^{N}{\rm e}^{ik_{e}x} +d_{1}\phi_{4}^{N}{\rm e}^{-ik_{h}x},\,x>L_{\rm S}&
\end{cases}\\
\Psi_{2}(x)&=
\begin{cases}
      \phi_{4}^{N}{\rm e}^{-ik_{h}x}+a_{2}\phi_{2}^{N}{\rm e}^{-ik_{e}x}+b_{2}\phi_{3}^{N}{\rm e}^{ik_{h}x},\, x<0& \\
             \sum_{i}q_{i}\phi_{i}^{S}{\rm e}^{ik_{i}^{S}x},\,0<x<L_{\rm S}& \\
   c_{2}\phi_{4}^{N}{\rm e}^{-ik_{h}x} +d_{2}\phi_{1}^{N}{\rm e}^{ik_{e}x},\,x>L_{\rm S}&
\end{cases}\\
\Psi_{3}(x)&=
\begin{cases}
 c_{3}\phi_{2}^{N}{\rm e}^{-ik_{e}x} +d_{3}\phi_{3}^{N}{\rm e}^{ik_{h}x},\, x<0& \\
     \sum_{i}r_{i}\phi_{i}^{S}{\rm e}^{ik_{i}^{S}x},\,0<x<L_{\rm S}& \\
 \phi_{2}^{N}{\rm e}^{-ik_{e}x}+a_{3}\phi_{4}^{N}{\rm e}^{-ik_{h}x}+b_{3}\phi_{1}^{N}{\rm e}^{ik_{e}x},\,x>L_{\rm S}&
\end{cases}\\
\Psi_{4}(x)&=
\begin{cases}
 c_{4}\phi_{3}^{N}{\rm e}^{ik_{h}x} +d_{4}\phi_{2}^{N}{\rm e}^{-ik_{e}x},\, x<0& \\
      \sum_{i}s_{i}\phi_{i}^{S}\,{\rm e}^{ik_{i}^{S}x},\,0<x<L_{\rm S}& \\
 \phi_{3}^{N}{\rm e}^{ik_{h}x}+a_{4}\phi_{1}^{N}{\rm e}^{ik_{e}x}+b_{4}\phi_{4}^{N}{\rm e}^{-ik_{h}x},\,x>L_{\rm S}\,,&
\end{cases}
\end{split}
\end{equation}
while the conjugated processes are
\begin{equation}
\label{eqnsn2}
\begin{split}
\tilde{\Psi}_{1}(x)&=
\begin{cases}
      \tilde{\phi}_{1}^{N}{\rm e}^{ik_{e}x}+\tilde{a}_{1}\tilde{\phi}_{3}^{N}{\rm e}^{ik_{h}x}+\tilde{b}_{1}\tilde{\phi}_{2}^{N}{\rm e}^{-ik_{e}x},\,x<0& \\
      \sum_{i}\tilde{p}_{i}\tilde{\phi}_{i}^{S}{\rm e}^{ik_{i}^{S}x},\,0<x<L_{\rm S}& \\
   \tilde{c}_{1}\tilde{\phi}_{1}^{N}{\rm e}^{ik_{e}x} +\tilde{d}_{1}\tilde{\phi}_{4}^{N}{\rm e}^{-ik_{h}x},\,x>L_{\rm S}\,,
\end{cases}\\
\tilde{\Psi}_{2}(x)&=
\begin{cases}
      \tilde{\phi}_{4}^{N}{\rm e}^{-ik_{h}x}+\tilde{a}_{2}\tilde{\phi}_{2}^{N}{\rm e}^{-ik_{e}x}+\tilde{b}_{2}\tilde{\phi}_{3}^{N}{\rm e}^{ik_{h}x},\, x<0& \\
            \sum_{i}\tilde{q}_{i}\tilde{\phi}_{i}^{S}{\rm e}^{ik_{i}^{S}x},\,0<x<L_{\rm S}& \\
   \tilde{c}_{2}\tilde{\phi}_{4}^{N}{\rm e}^{-ik_{h}x} +\tilde{d}_{2}\tilde{\phi}_{1}^{N}{\rm e}^{ik_{e}x},\,x>L_{\rm S}&
\end{cases}\\
\tilde{\Psi}_{3}(x)&=
\begin{cases}
 \tilde{c}_{3}\tilde{\phi}_{2}^{N}{\rm e}^{-ik_{e}x} +\tilde{d}_{3}\tilde{\phi}_{3}^{N}{\rm e}^{ik_{h}x},\, x<0& \\
      \sum_{i}\tilde{r}_{i}\tilde{\phi}_{i}^{S}{\rm e}^{ik_{i}^{S}x},\,0<x<L_{\rm S}& \\
 \tilde{\phi}_{2}^{N}{\rm e}^{-ik_{e}x}+\tilde{a}_{3}\tilde{\phi}_{4}^{N}{\rm e}^{-ik_{h}x}+\tilde{b}_{3}\tilde{\phi}_{1}^{N}{\rm e}^{ik_{e}x},\,x>L_{\rm S}&
\end{cases}\\
\tilde{\Psi}_{4}(x)&=
\begin{cases}
 \tilde{c}_{4}\tilde{\phi}_{3}^{N}{\rm e}^{ik_{h}x} +\tilde{d}_{4}\tilde{\phi}_{2}^{N}{\rm e}^{-ik_{e}x},\, x<0 &\\
      \sum_{i}\tilde{s}_{i}\tilde{\phi}_{i}^{S}{\rm e}^{ik_{i}^{S}x},\,0<x<L_{\rm S}& \\
 \tilde{\phi}_{3}^{N}{\rm e}^{ik_{h}x}+\tilde{a}_{4}\tilde{\phi}_{1}^{N}{\rm e}^{ik_{e}x}+\tilde{b}_{4}\tilde{\phi}_{4}^{N}{\rm e}^{-ik_{h}x},\,x>L_{\rm S}&
\end{cases}
\end{split}
\end{equation}
where $i=1,2,3,4$, $ k_{1,2}^{S}=\pm k_{e}^{S}$, $ k_{3,4}^{S}=\pm k_{h}^{S}$ 
and the spinors $\phi_{i}^{N,S}$ and $\tilde{\phi}_{i}^{N,S}$ are given by Eqs.\,(\ref{phiNS}) 
and (\ref{tildephiNS}), respectively. 

Notice that, as for NS junctions, in this part we have written the general form of the scattering wave functions in the N and S regions. 
The S region is in general formed by a linear combination of four elements given in terms of the amplitudes $p_{i}$, $q_{i}$, $r_{i}$ and $s_{i}$. Their meaning is as follows: $p_{1}$ represent electron transmission from left N to S, $p_{2}$ the normal reflection at the right SN interface, $p_{3}$ the Andreev reflection at the SN interface, and $p_{4}$ the Andreev reflection at the left NS interface. Similar ideas apply to the amplitudes $q_{i}$, $r_{i}$ and $s_{i}$.
The amplitudes of all these processes are calculated after matching the wave-functions at the left NS and right SN interfaces.
Due to  helicity conservation in the 2DTI edge states we can again directly obtain 
$b_{i}=d_{i}=\tilde{b}_{i}=\tilde{d}_{i}=0$ and $p_{2,4}=s_{2,4}=r_{1,3}=q_{1,3}=0$. Similar relations hold for the conjugated processes.

\subsection{Green's function and pairing amplitudes}
\subsubsection{Normal regions}
\label{GNSN_N}
Finding $\Psi_{i}$ and $\tilde{\Psi}_{i}$, given by Eqs.\,(\ref{eqnsn1}) and (\ref{eqnsn2}), allows us to construct the retarded Green's functions using Eq.\,(\ref{GFUNCTION}) in the main text.
In the two normal regions the Green's functions and pairing amplitudes acquire the same form as in the NS junction, with the only difference that the Andreev reflection amplitude becomes dependent on the length $L_{\rm S}$ of the S region and reads  $a_1(\omega,L_{\rm S})= a_{2}(\omega,L_{\rm S})={\rm sin}[i\kappa(\omega)L]/{\rm sin}[i\kappa(\omega)L-\eta(\omega)]$.  
Notice that $a_{2}(\omega,L_{\rm S})\rightarrow a_{2}(\omega)=v/u$ for $L_{\rm S}\rightarrow\infty$, as expected.

\subsubsection{Superconducting region}
\label{GNSN_S}
Due to the finite length of the S region in NSN junctions we find quite different results here compared to the NS junction. In the S region the retarded Green's function contains the components
\begin{widetext}

\begin{equation}
\begin{split}
G^{r}_{ee,\uparrow\uparrow}(x,x',\omega)&={\rm e}^{i(k_{h}^{S}x-k_{e}^{S}x')}\gamma_{2}+
{\rm e}^{i(k_{e}^{S}x-k_{h}^{S}x')}\gamma_{3}+\theta(x-x')\Big[{\rm e}^{ik_{e}^{S}(x-x')}\gamma_{1}+{\rm e}^{ik_{h}^{S}(x-x')}\gamma_{4}\Big]\\
&-\theta(x'-x)\Big[{\rm e}^{ik_{e}^{S}(x-x')}\gamma_{2}+{\rm e}^{ik_{h}^{S}(x-x')}\gamma_{3}\Big]\,,\\
G^{r}_{ee,\downarrow\downarrow}(x,x',\omega)&={\rm e}^{i(k_{e}^{S}x'-k_{h}^{S}x)}\gamma_{3}+
{\rm e}^{i(k_{h}^{S}x'-k_{e}^{S}x)}\gamma_{2}-\theta(x-x')\Big[{\rm e}^{-ik_{e}^{S}(x-x')}\gamma_{2}+{\rm e}^{-ik_{h}^{S}(x-x')}\gamma_{3}\Big]\\
&+\theta(x'-x)\Big[{\rm e}^{-ik_{e}^{S}(x-x')}\gamma_{1}+{\rm e}^{-ik_{h}^{S}(x-x')}\gamma_{4}\Big]\,,\\
G^{r}_{hh,\downarrow\downarrow}(x,x',\omega)&={\rm e}^{i(k_{h}^{S}x-k_{e}^{S}x')}\gamma_{2}+
{\rm e}^{i(k_{e}^{S}x-k_{h}^{S}x')}\gamma_{3}-\theta(x-x')\Big[{\rm e}^{ik_{h}^{S}(x-x')}\gamma_{2}+{\rm e}^{ik_{e}^{S}(x-x')}\gamma_{3}\Big]\\
&+\theta(x'-x)\Big[{\rm e}^{ik_{h}^{S}(x-x')}\gamma_{1}+{\rm e}^{ik_{e}^{S}(x-x')}\gamma_{4}\Big]\,,\\
G^{r}_{hh,\uparrow\uparrow}(x,x',\omega)&={\rm e}^{i(k_{e}^{S}x'-k_{h}^{S}x)}\gamma_{3}+
{\rm e}^{i(k_{h}^{S}x'-k_{e}^{S}x)}\gamma_{2}+\theta(x-x')\Big[{\rm e}^{-ik_{h}^{S}(x-x')}\gamma_{1}+{\rm e}^{-ik_{e}^{S}(x-x')}\gamma_{4}\Big]\\
&-\theta(x'-x)\Big[{\rm e}^{-ik_{h}^{S}(x-x')}\gamma_{2}+{\rm e}^{-ik_{e}^{S}(x-x')}\gamma_{3}\Big]\,,\\
G_{eh,\uparrow\downarrow}^{r}(x,x',\omega)&=-{\rm e}^{i(k_{e}^{S}x-k_{h}^{S}x')}\beta_{2}-
{\rm e}^{i(k_{h}^{S}x-k_{e}^{S}x')}\beta_{3}+{\rm e}^{i k_{e}^{S}(x-x')}\Big[\theta(x-x')\beta_{2}+\theta(x'-x)\beta_{3}\Big]\\
&+{\rm e}^{i k_{h}^{S}(x-x')}\Big[\theta(x-x')\beta_{3}+\theta(x'-x)\beta_{2}\Big]\,,\\
G_{eh,\downarrow\uparrow}^{r}(x,x',\omega)&={\rm e}^{i(k_{h}^{S}x'-k_{e}^{S}x)}\beta_{1}+
{\rm e}^{i(k_{e}^{S}x'-k_{h}^{S}x)}\beta_{4}+{\rm e}^{-i k_{e}^{S}(x-x')}\Big[\theta(x-x')\beta_{3}+\theta(x'-x)\beta_{2}\Big]\\
&+{\rm e}^{-i k_{h}^{S}(x-x')}\Big[\theta(x-x')\beta_{2}+\theta(x'-x)\beta_{3}\Big]\,,\\
G_{he,\downarrow\uparrow}^{r}(x,x',\omega)&={\rm e}^{i(k_{h}^{S}x-k_{e}^{S}x')}\beta_{1}+
{\rm e}^{i(k_{e}^{S}x-k_{h}^{S}x')}\beta_{4}+{\rm e}^{i k_{e}^{S}(x-x')}\Big[\theta(x-x')\beta_{2}+\theta(x'-x)\beta_{3}\Big]\\
&+{\rm e}^{i k_{h}^{S}(x-x')}\Big[\theta(x-x')\beta_{3}+\theta(x'-x)\beta_{2}\Big]\,,\\
G_{he,\uparrow\downarrow}^{r}(x,x',\omega)&=-{\rm e}^{i(k_{e}^{S}x'-k_{h}^{S}x)}\beta_{2}-
{\rm e}^{i(k_{h}^{S}x'-k_{e}^{S}x)}\beta_{3}+{\rm e}^{-i k_{e}^{S}(x-x')}\Big[\theta(x-x')\beta_{3}+\theta(x'-x)\beta_{2}\Big]\\
&+{\rm e}^{-i k_{h}^{S}(x-x')}\Big[\theta(x-x')\beta_{2}+\theta(x'-x)\beta_{3}\Big]\,,
\end{split}
\end{equation}
where we have defined
\begin{equation}
\begin{split}
\gamma_{1}&=\frac{1}{iv_{f}}\frac{u^{4}}{(u^{2}-v^{2})[u^{2}-{\rm e}^{i(k_{e}^{S}-k_{h}^{S})L}v^{2}]}\,,\quad
\gamma_{2}=\frac{1}{iv_{f}}\frac{u^{2}v^{2}}{(u^{2}-v^{2})[-u^{2}{\rm e}^{i(k_{h}^{S}-k_{e}^{S})L}+v^{2}]}\,,\\
\gamma_{3}&=\frac{1}{iv_{f}}\frac{u^{2}v^{2}}{(u^{2}-v^{2})[-u^{2}+{\rm e}^{i(k_{e}^{S}-k_{h}^{S})L}v^{2}]}\,,\quad
\gamma_{4}=\frac{1}{iv_{f}}\frac{v^{4}}{(u^{2}-v^{2})[u^{2}{\rm e}^{i(k_{h}^{S}-k_{e}^{S})L}-v^{2}]}\,,\\
\beta_{1}&=
\frac{1}{iv_{f}}\frac{u^{3}v}{(u^{2}-v^{2})[-u^{2}{\rm e}^{i(k_{h}^{S}-k_{e}^{S})L}+v^{2}]}\,,\quad
\beta_{2}=\frac{1}{iv_{f}}\frac{u^{3}v}{(u^{2}-v^{2})[u^{2}-{\rm e}^{i(k_{e}^{S}-k_{h}^{S})L}v^{2}]}\,,\\
\beta_{3}&=\frac{1}{iv_{f}}\frac{u v^{3}}{(u^{2}-v^{2})[u^{2}{\rm e}^{i(k_{h}^{S}-k_{e}^{S})L}-v^{2}]}\,,\quad
\beta_{4}=\frac{1}{iv_{f}}\frac{u v^{3}}{(u^{2}-v^{2})[-u^{2}+{\rm e}^{i(k_{e}^{S}-k_{h}^{S})L}v^{2}]}\,.
\end{split}
\end{equation}
The regular electron-electron part of $G^{r}$ gives the LDOS in the S region which reads \begin{equation}
\label{LDOS_NSN1}
\rho_{S}(x,\omega)=
-\frac{1}{\pi v_{F}}{\rm Im}\Big\{\frac{\gamma_{1}}{i}
\Big[\frac{v^{2}}{u^{2}}{\rm e}^{2ik(\omega)L}\Big(1+\frac{v^{2}}{u^{2}}-2{\rm e}^{-2ik(\omega)x}\Big)+\Big(1+\frac{v^{2}}{u^{2}}-2\frac{v^{2}}{u^{2}}{\rm e}^{2ik(\omega)x} \Big)\Big]  \Big\}\,.
\end{equation}
For energies within $\Delta$ this expression reduces to 
\begin{equation}
\label{LDOS_NSN2}
\rho_{\rm S}(x,|\omega|<\Delta)=\frac{\omega^{2}(1-{\rm e}^{-2\kappa(\omega)L})^{2}+\Delta^{2}(1+{\rm e}^{-2\kappa(\omega)L})\Big[{\rm e}^{-2\kappa(\omega)L}\Big({\rm e}^{2\kappa(\omega)x}-\frac{\omega^{2}}{\Delta^{2}}\Big) +\Big({\rm e}^{-2\kappa(\omega)x}-\frac{\omega^{2}}{\Delta^{2}}\Big)\Big]}{\pi v_{F}[\Delta^{2}(1+{\rm e}^{-2\kappa(\omega)L})^{2}-4\omega^{2}{\rm e}^{-2\kappa(\omega)L}]}\,.
\end{equation}
At $\omega=0$, this expression can be simplified even further and we obtain 
\begin{equation}
\label{LDOS_NSN3}
\rho_{\rm S}(x,0)=\frac{1}{\pi v_{F}}\frac{{\rm e}^{-2L/\xi}{\rm e}^{2x/\xi}+{\rm e}^{-2x/\xi}}{1+{\rm e}^{-2L/\xi}}\,,
\end{equation}
where we have used $\xi = v_F/\Delta$. This is the expression for the LDOS given in Eq.\,(\ref{LDOS_NSN3x}) in the main text.

For the pairing amplitudes, limiting ourselves to energies within the gap, we arrive at
\begin{equation}
\label{f_NSN_S2}
\begin{split}
f_{0,{\rm S}}^{r}(x,x',\omega)&=
-\beta_{2}{\rm e}^{-\kappa(\omega)(x+x')}\Big[{\rm e}^{-ik_{\mu}(x-x')-2i\eta(\omega)} +{\rm e}^{ik_{\mu}(x-x')} \Big]+\beta_{1}{\rm e}^{\kappa(\omega)(x+x')}\Big[{\rm e}^{-ik_{\mu}(x-x')} +{\rm e}^{ik_{\mu}(x-x')-2i\eta(\omega)} \Big]\\
&+2{\rm cos}[k_{\mu}(x-x')]\Big[\beta_{2}{\rm e}^{-\kappa(\omega)|x-x'|}-\beta_{1}{\rm e}^{-2i\eta(\omega)}{\rm e}^{\kappa(\omega)|x-x'|}\Big]\,,\\
f_{3,{\rm S}}^{r}(x,x',\omega)&=
-\beta_{2}{\rm e}^{-\kappa(\omega)(x+x')}
\Big[{\rm e}^{ik_{\mu}(x-x')}-{\rm e}^{-ik_{\mu}(x-x')-2i\eta(\omega)}  \Big]+
\beta_{1}{\rm e}^{\kappa(\omega)(x+x')}\Big[{\rm e}^{ik_{\mu}(x-x')-2i\eta(\omega)} -{\rm e}^{-ik_{\mu}(x-x')} \Big]\\
&+2i{\rm sin}[k_{\mu}(x-x')]\Big[\beta_{2}{\rm e}^{-\kappa(\omega)|x-x'|}-\beta_{1}{\rm e}^{-2i\eta(\omega)}{\rm e}^{\kappa(\omega)|x-x'|}\Big]\nonumber
\,,\\
\end{split}
\end{equation}
and $f^r_{1,{\rm S}} = f^r_{2,{\rm S}} = 0$, where we have used $\gamma_{4,2}=-{\rm e}^{-2i\eta(\omega)}\gamma_{2,1}$ and $\beta_{4,3}=-{\rm e}^{-2i\eta(\omega)}\beta_{2,1}$, 
with 
\begin{equation}
\beta_{1}=-\frac{iZ(\omega){\rm e}^{i\eta(\omega)-\kappa(\omega)L_{\rm S}}}{4{\rm sin}[i\kappa(\omega)L_{\rm S}+\eta(\omega)]}\nonumber\,, 
\end{equation}
$\beta_{2}=-\beta_{1}{\rm e}^{2\kappa(\omega)L_{\rm S}}$, $\gamma_{1}={\rm e}^{i\eta(\omega)}\beta_{2}$,  and $\gamma_{2}={\rm e}^{-i\eta(\omega)}\beta_{1}$.
From these and the related advanced expressions we can extract the even- and odd-frequency components which reads
\begin{equation}
\label{f_NSN_S3}
\begin{split}
f_{0,{\rm S}}^{r,{\rm E}}(x,x',\omega)&=C_{xx'}
\Big[{\rm e}^{-\kappa(\omega)(x+x')}\beta_{42}^{-}+{\rm e}^{\kappa(\omega)(x+x')}\beta_{13}^{-}+2\beta_{2}{\rm e}^{-\kappa(\omega)|x-x'|}+2\beta_{3}{\rm e}^{\kappa(\omega)|x-x'|}\Big]\,,\\
f_{0,{\rm S}}^{r,{\rm O}}(x,x',\omega)&=-iS_{xx'}
\Big[{\rm e}^{-\kappa(\omega)(x+x')}\beta_{42}^{+}+{\rm e}^{\kappa(\omega)(x+x')}\beta_{13}^{+}\,,\\
f_{3,{\rm S}}^{r,{\rm E}}(x,x',\omega)&=iS_{xx'}
\Big[{\rm e}^{-\kappa(\omega)(x+x')}\beta_{42}^{-}+{\rm e}^{\kappa(\omega)(x+x')}\beta_{13}^{-}+2\beta_{2}{\rm e}^{-\kappa(\omega)|x-x'|}+2\beta_{3}{\rm e}^{\kappa(\omega)|x-x'|}\Big]\,,\\
f_{3,{\rm S}}^{r,{\rm O}}(x,x',\omega)&=-C_{xx'}
\Big[{\rm e}^{-\kappa(\omega)(x+x')}\beta_{42}^{+}+{\rm e}^{\kappa(\omega)(x+x')}\beta_{13}^{+}\Big]\,,
\end{split}
\end{equation}
where $\beta_{42(13)}^{\pm}=\beta_{4(1)}\pm\beta_{2(3)}$.
These pairing amplitudes in S can be composed into bulk (B) and interface (I) contributions. Bulk contributions we designate terms which are independent of the average distance from either interface, i.e.~with an overall $|x-x'|$ spatial dependence. The remaining terms we label interface contributions, as they all have a decay length $1/[2\kappa(\omega)]$ from the interface.
With this division we directly arrive at the results given in Eqs.~(\ref{f_NSN_S3_main1})-(\ref{f_NSN_S3_main2}) in the main text. 
\end{widetext}

\section{SNS junction}
\label{SNSApp}
Finally we consider SNS junctions, where the normal region has a finite length $L_{\rm N}$, while the S regions are semi-infinite such that
\begin{equation}
\Delta(x)=\begin{cases}
   \Delta \,,  & x<0, \\
    0\,, & x<0<L_{\rm N},\\
   \Delta\, {\rm e}^{i\phi} \,,  & x>0, 
\end{cases}
\end{equation}
where $\phi$ is the superconducting phase difference across the junction.

\subsection{Scattering processes}
In this case the processes read
\begin{equation}
\label{eqsns1}
\begin{split}
\Psi_{1}(x)&=
\begin{cases}
      \phi_{1}^{S_{L}}\,{\rm e}^{ik^{S}_{e}x}+a_{1}\,\phi_{3}^{S_{L}}\,{\rm e}^{ik^{S}_{h}x}+b_{1}\,\phi_{2}^{S_{L}}\,{\rm e}^{-ik_{e}^{S}x},\, x<0& \\
 \sum_{i}p_{i}\,\phi_{i}^{N}\,{\rm e}^{ik_{i}x},\, 0<x<L_{\rm N}&\\
    c_{1}\,\phi_{1}^{S_{R}}\,{\rm e}^{ik_{e}^{S}x} +d_{1}\,\phi_{4}^{S_{R}}\,{\rm e}^{-ik_{h}^{S}x},\,x>L_{\rm N}&
\end{cases}\\
\Psi_{2}(x)&=
\begin{cases}
      \phi_{4}^{S_{L}}{\rm e}^{-ik_{h}^{S}x}+a_{2}\phi_{2}^{S_{L}}{\rm e}^{-ik^{S}_{e}x}+b_{2}\phi_{3}^{S_{L}}{\rm e}^{ik_{h}^{S}x},\, x<0& \\
  \sum_{i}q_{i}\phi_{i}^{N}{\rm e}^{ik_{i}x},\, 0<x<L_{\rm N}&\\
   c_{2}\phi_{4}^{S_{R}}{\rm e}^{-ik^{S}_{h}x} +d_{2}\phi_{1}^{S_{R}}{\rm e}^{ik^{S}_{e}x},\,x>L_{\rm N}&
\end{cases}\\
\Psi_{3}(x)&=
\begin{cases}
 c_{3}\phi_{2}^{S_{L}}{\rm e}^{-ik_{e}^{S}x} +d_{3}\phi_{3}^{S_{L}}{\rm e}^{ik^{S}_{h}x},\, x<0& \\
  \sum_{i}r_{i}\phi_{i}^{N}{\rm e}^{ik_{i}x},\, 0<x<L_{\rm N}&\\
 \phi_{2}^{S_{R}}{\rm e}^{-ik^{S}_{e}x}+a_{3}\phi_{4}^{S_{R}}{\rm e}^{-ik^{S}_{h}x}+b_{3}\phi_{1}^{S_{R}}{\rm e}^{ik^{S}_{e}x},\,x>L_{\rm N}&
\end{cases}\\
\Psi_{4}(x)&=
\begin{cases}
 c_{4}\phi_{3}^{S_{L}}{\rm e}^{ik^{S}_{h}x} +d_{4}\phi_{2}^{S_{L}}{\rm e}^{-ik^{S}_{e}x},\, x<0& \\
  \sum_{i}s_{i}\phi_{i}^{N}{\rm e}^{ik_{i}x},\, 0<x<L_{\rm N}&\\
 \phi_{3}^{S_{R}}{\rm e}^{ik^{S}_{h}x}+a_{4}\phi_{1}^{S_{R}}{\rm e}^{ik_{e}^{S}x}+b_{4}\phi_{4}^{S_{R}}{\rm e}^{-ik^{S}_{h}x},\,x>L_{\rm N}&
\end{cases}
\end{split}
\end{equation}

\begin{equation}
\label{eqsns2}
\begin{split}
\tilde{\Psi}_{1}(x)&=
\begin{cases}
      \tilde{\phi}_{1}^{S_{L}}{\rm e}^{ik^{S}_{e}x}+\tilde{a}_{1}\tilde{\phi}_{3}^{S_{L}}\,{\rm e}^{ik^{S}_{h}x}+\tilde{b}_{1}\tilde{\phi}_{2}^{S_{L}}{\rm e}^{-ik^{S}_{e}x},\, x<0& \\
        \sum_{i}\tilde{p}_{i}\tilde{\phi}_{i}^{N}{\rm e}^{ik_{i}x},\, 0<x<L_{\rm N}&\\
   \tilde{c}_{1}\tilde{\phi}_{1}^{S_{R}}{\rm e}^{ik_{e}^{S}x} +\tilde{d}_{1}\tilde{\phi}_{4}^{S_{R}}{\rm e}^{-ik^{S}_{h}x},\,x>L_{\rm N}&
\end{cases}\\
\tilde{\Psi}_{2}(x)&=
\begin{cases}
      \tilde{\phi}_{4}^{S_{L}}{\rm e}^{-ik^{S}_{h}x}+\tilde{a}_{2}\tilde{\phi}_{2}^{S_{L}}{\rm e}^{-ik_{e}^{S}x}+\tilde{b}_{2}\tilde{\phi}_{3}^{S_{L}}{\rm e}^{ik^{S}_{h}x},\, x<0& \\
              \sum_{i}\tilde{q}_{i}\tilde{\phi}_{i}^{N}{\rm e}^{ik_{i}x},\, 0<x<L_{\rm N}&\\
   \tilde{c}_{2}\tilde{\phi}_{4}^{S_{R}}{\rm e}^{-ik^{S}_{h}x} +\tilde{d}_{2}\tilde{\phi}_{1}^{S_{R}}{\rm e}^{ik^{S}_{e}x},\,x>L_{\rm N}&
\end{cases}\\
\tilde{\Psi}_{3}(x)&=
\begin{cases}
 \tilde{c}_{3}\tilde{\phi}_{2}^{S_{L}}{\rm e}^{-ik^{S}_{e}x} +\tilde{d}_{3}\tilde{\phi}_{3}^{S_{L}}{\rm e}^{ik^{S}_{h}x},\, x<0& \\
         \sum_{i}\tilde{r}_{i}\tilde{\phi}_{i}^{N}{\rm e}^{ik_{i}x},\, 0<x<L_{\rm N}&\\
 \tilde{\phi}_{2}^{S_{R}}{\rm e}^{-ik^{S}_{e}x}+\tilde{a}_{3}\tilde{\phi}_{4}^{S_{R}}{\rm e}^{-ik^{S}_{h}x}+\tilde{b}_{3}\tilde{\phi}_{1}^{S_{R}}{\rm e}^{ik^{S}_{e}x},\,x>L_{\rm N}&
\end{cases}\\
\tilde{\Psi}_{4}(x)&=
\begin{cases}
 \tilde{c}_{4}\tilde{\phi}_{3}^{S_{L}}{\rm e}^{ik_{h}^{S}x} +\tilde{d}_{4}\tilde{\phi}_{2}^{S_{L}}{\rm e}^{-ik_{e}^{S}x},\, x<0& \\
\sum_{i}\tilde{s}_{i}\tilde{\phi}_{i}^{N}{\rm e}^{ik_{i}x},\, 0<x<L_{\rm N}&\\
 \tilde{\phi}_{3}^{S_{R}}{\rm e}^{ik_{h}^{S}x}+\tilde{a}_{4}\tilde{\phi}_{1}^{S_{R}}{\rm e}^{ik_{e}^{S}x}+\tilde{b}_{4}\tilde{\phi}_{4}^{S_{R}}{\rm e}^{-ik^{S}_{h}x}
,\,x>L_{\rm N}&
 \end{cases}
\end{split}
\end{equation}
where $i=1,2,3,4$ and the spinors $\phi_{i}^{N}$, $\tilde{\phi}_{i}^{N}$, 
$\phi_{i}^{S_{L}}$, $\tilde{\phi}_{i}^{S_{L}}$ acquire the same form as in the NS junction, 
given by Eqs.\,(\ref{phiNS}) and (\ref{tildephiNS}). In the right superconductor, 
the wave functions of $H_{\rm BdG}$ read
\begin{equation}
 \begin{split}
 \phi_{1}^{S_{R}}&=\begin{pmatrix}
u\,{\rm e}^{i\phi/2},
0,
v\,{\rm e}^{-i\phi/2},
0
 \end{pmatrix}^{T}\\
\phi_{2}^{S_{R}}&=\begin{pmatrix}
0,
u\,{\rm e}^{i\phi/2},
0,
v\,{\rm e}^{-i\phi/2}
 \end{pmatrix}^{T}\\
\phi_{3}^{S_{R}}&=\begin{pmatrix}
v\,{\rm e}^{i\phi/2},
0,
u\,{\rm e}^{-i\phi/2},
0
 \end{pmatrix}^{T}\\
\phi_{4}^{S_{R}}&=\begin{pmatrix}
0,
v\,{\rm e}^{i\phi/2},
0,
u\,{\rm e}^{-i\phi/2}
 \end{pmatrix}^{T},
\end{split}
\end{equation}
while the corresponding conjugated wave functions are
\begin{equation}
\begin{split}
\tilde{\phi}_{1}^{S_{R}}&=
\begin{pmatrix}
0,
u\,{\rm e}^{-i\phi/2},
0,
v\,{\rm e}^{i\phi/2}
\end{pmatrix}^{T}\\
\tilde{\phi}_{2}^{S_{R}}&=
\begin{pmatrix}
u\,{\rm e}^{-i\phi/2},
0,
v\,{\rm e}^{i\phi/2},
0
\end{pmatrix}^{T}\\
\tilde{\phi}_{3}^{S_{R}}&=
\begin{pmatrix}
0,
v\,{\rm e}^{-i\phi/2},
0,
u\,{\rm e}^{i\phi/2}
\end{pmatrix}^{T}\\
\tilde{\phi}_{4}^{S_{R}}&=
\begin{pmatrix}
v\,{\rm e}^{-i\phi/2},
0,
u\,{\rm e}^{i\phi/2},
0
\end{pmatrix}^{T}\,.
\end{split}
\end{equation}
 As before,  helicity conservation directly imposes $b_{i}=d_{i}=\tilde{b}_{i}=\tilde{d}_{i}=0$ 
 and also $q_{1,3}=r_{1,3}=p_{2,4}=s_{2,4}=\tilde{q}_{1,3}=\tilde{r}_{1,3}=\tilde{p}_{2,4}=\tilde{s}_{2,4}=0$.

\subsection{Green's functions and pairing amplitudes}
\subsubsection{Normal region}
We construct the retarded Green's function from the scattering states in the same was as for SN 
and NSN junctions. In the normal region N the regular and anomalous Green's functions  become
\begin{equation}
\label{SNS_G_N}
\begin{split}
G_{ee}^{r}(x,x',\omega)&=
\begin{pmatrix}
{\rm e}^{ik_{e}(x-x')}
M_{1}(x,x')&0\\
0& {\rm e}^{-ik_{e}(x-x')}M_{2}(x,x')
\end{pmatrix}\,,\\
G_{hh}^{r}(x,x',\omega)&=
\begin{pmatrix}
{\rm e}^{ik_{h}(x-x')}M_{1}(x',x)&0\\
0& {\rm e}^{-ik_{h}(x-x')}M_{2}(x',x)
\end{pmatrix}\,,\\
G_{eh}^{r}(x,x',\omega)&=
\begin{pmatrix}
{\rm e}^{i(k_{e}x-k_{h}x')}\,m_{5}&0\\
0&{\rm e}^{-i(k_{e}x-k_{h}x')}\,m_{6}
\end{pmatrix}\,,\\
G_{he}^{r}(x,x',\omega)&=
\begin{pmatrix}
{\rm e}^{i(k_{h}x-k_{e}x')}\,m_{7}&0\\
0&{\rm e}^{-i(k_{h}x-k_{e}x')}\,m_{8}
\end{pmatrix}\,,
\end{split}
\end{equation}
where $M_{1}(x,x')=\theta(x-x')m_{1}+\theta(x'-x)m_{3}$,  
$M_{2}(x,x')=\theta(x-x')m_{2}+\theta(x'-x)m_{4}$, 
$m_{1}=\frac{u}{v}m_{5}, m_{2}=\frac{v}{u}m_{6}, m_{3}=\frac{v}{u}m_{7}, m_{4}=\frac{u}{v}m_{8}$, and
\begin{equation}
\label{mSNS}
\begin{split}
m_{5}&\equiv p_{1}\tilde{r}_{4}\alpha_{1}=\frac{1}{iv_{f}}\frac{uv}{u^{2}-v^{2}\,{\rm e}^{i(k_{e}-k_{h})L_{\rm N}-i\phi}}\,,\\
m_{6}&\equiv q_{2}\tilde{s}_{3}\alpha_{4}=\frac{1}{iv_{f}}\frac{uv}{u^{2}\,{\rm e}^{i(k_{h}-k_{e})L_{\rm N}-i\phi}-v^{2}}\,,\\
m_{7}&\equiv r_{1}\tilde{r}_{2}\alpha_{1}=\frac{1}{iv_{f}}\frac{uv}{u^{2}\,{\rm e}^{i(k_{h}-k_{e})L_{\rm N}+i\phi}-v^{2}}\,,\\
m_{8}&\equiv q_{4}\tilde{s}_{1}\alpha_{4}=\frac{1}{iv_{f}}\frac{uv}{u^{2}-v^{2}\,{\rm e}^{i(k_{e}-k_{h})L_{\rm N}+i\phi}}\,.
\end{split}
\end{equation}
From these equations, we observe that  it is enough to specify the form of $m_{5}$. 
Thus, for energies within the superconducting gap previous expressions reduce to\begin{equation}
\label{mSNSApp}
 \begin{split}
  m_{5}(\omega,L_{\rm N},\phi)&=-\frac{1}{2v_{F}}\frac{{\rm e}^{i(\phi/2- L_{\rm N}/\xi_{\omega})}}{{\rm sin}[\eta(\omega)-L_{\rm N}/\xi_{\omega}+\phi/2]}\,\\
  m_{6}(\omega,L_{\rm N},\phi)&=m_{5}(\omega,L_{\rm N},-\phi){\rm e}^{i(2L_{\rm N}/\xi_{\omega}+\phi)}\,,\\
  m^{*}_{7,8}(-\omega)&=m_{6,5}(\omega)\,,
 \end{split}
\end{equation}
which correspond to Eq.\,(\ref{m5}) presented in the main text.

Instead of deriving the full expression for the LDOS from the regular part of $G^{r}$, we here focus on the main peaks as that is much easier and, for this discussion, as enlightening.
In the N region the low-energy peaks in the LDOS are from the discrete ABSs, whose number depend on the ratio $L_{\rm N}/\xi$. There exist different ways to calculate these energy levels. One efficient way is to locate the poles (zeros of the denominator) of the Andreev reflection coefficients, $a_{1,2}$, which are
\begin{equation}
\label{ARCOEFFI}
a_{1}(\omega,L_{\rm N},\phi)=\frac{uv\big[{\rm e}^{i(k_{e}L_{\rm N}-\phi/2)}-{\rm e}^{i(k_{h}L_{\rm N}
+\phi/2)}\big]}{u^{2}{\rm e}^{i(k_{h}L_{\rm N}+\phi/2)}-v^{2}{\rm e}^{i(k_{e}L_{\rm N}-\phi/2)}}\,.\\
\end{equation}
For energies within $\Delta$, and using $u/v={\rm e}^{i\eta(\omega)}$, this expression can be written as
\begin{equation}
a_{1}(\omega,L_{\rm N},\phi)=\frac{{\rm sin}[L_{\rm N}/\xi_{\omega}-\phi/2]}{{\rm sin}[\eta(\omega)-L_{\rm N}/\xi_{\omega}+\phi/2]}
\end{equation}
and also $a_{2}(\omega,L_{\rm N},\phi)=a_{1}(\omega,L_{\rm N},-\phi)$. With these expressions, the poles of $a_{1,2}$ can be shown to give rise to the following condition for the ABSs, which is also given in the main text:
\begin{equation}
\label{ABSsApp}
2\eta(\omega)-2L_{\rm N}/\xi_{\omega}\pm\phi=2\pi n\,,\quad n=0,1,\ldots.
\end{equation}
For a short junction, $L_{\rm N}\ll\xi$, the condition for ABSs reduces to $2\eta(\omega)\pm\phi=2\pi n$, with two Andreev levels at energies
\begin{equation}
\label{ABSsAppS}
\omega_{\pm}(\phi)=\pm\Delta{\rm cos}(\phi/2)\,,
\end{equation}
while for a long junction, $L_{\rm N}\gg\xi$ we obtain
\begin{equation}
\label{ABSsAppL}
\omega_{\pm}^{n}(\phi)=\frac{v_{F}}{2L_{\rm N}}\Big[2\pi\Big(n+\frac{1}{2}\Big)\pm\phi\Big]\,.
\end{equation}
\begin{figure}[!ht]
\centering
\includegraphics[width=.49\textwidth]{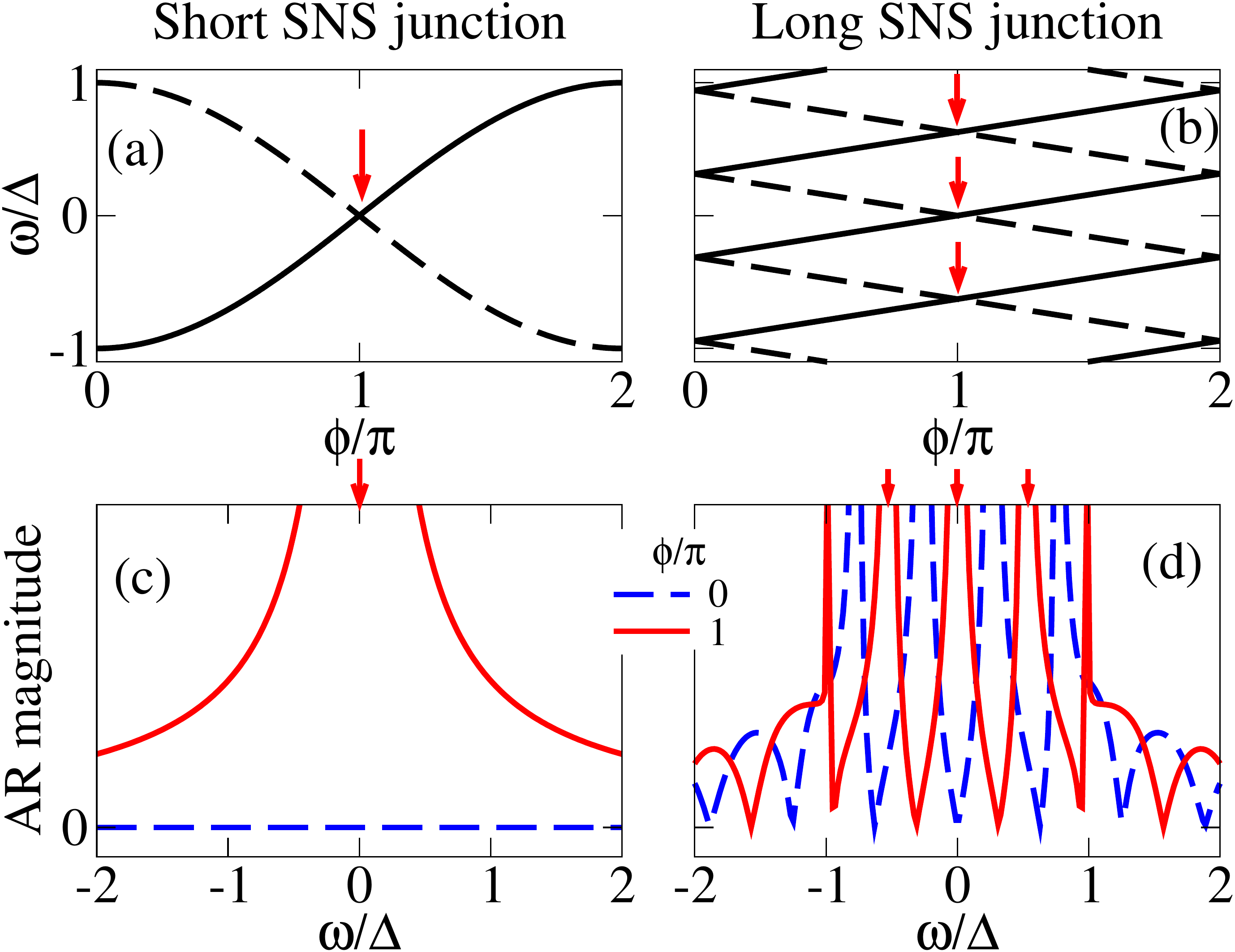} 
\caption{(Color online) ABSs in short (a) and long (b) SNS junctions at the edge of a 2DTI and the Andreev reflection coefficient $a_{1}$ at $\phi=0,\pi$ for short (c) and long (d) junctions. Parameters: $L_{\rm N}=5\xi$.
}
\label{ABSsA}
\end{figure}
Notice that the Andreev bound states given by Eqs.\,(\ref{ABSsAppS}) and \,(\ref{ABSsAppL}), 
and plotted in Fig.\,\ref{ABSsA}(a-b), develop crossings (indicated by red arrows) around zero energy 
$\omega=0$ at $\pi(2n-1)$ for $n=1,2,3,\ldots$,  which are protected by time-reversal symmetry. 

The Andreev reflection magnitude $|a_{1}(\omega,L_{\rm N},\phi)|$ at $\phi=0,\pi$ is also plotted in Fig.\,\ref{ABSsA}(c-d) for short and long junctions. Observe that it develops resonances at the energies of the protected crossings. In particular, when $\phi=\pi$ a single zero-energy peak emerges in short junctions as shown in Fig.\,\ref{ABSsA}(c). In longer junctions the number of such resonances (that correspond to ABSs) increases, being proportional to $L_{\rm N}/\xi$, and the zero-energy peak coexists with additional peaks as we observe in Fig.\,\ref{ABSsA}(d). 
Remarkably, the same discussion was performed in the main text but there instead the ABSs energies was derived from the pairing amplitudes.

In terms of the pairing amplitudes in the N region we obtain after some algebra
\begin{equation}
\label{SNS_Nxxx}
\begin{split}
f_{0,{\rm N}}^{r}(x,x',\omega)&=\frac{{\rm e}^{i(k_{e}x-k_{h}x')}\,m_{5}+{\rm e}^{-i(k_{e}x-k_{h}x')}\,m_{6}}{2}\,,\\
f_{3,{\rm N}}^{r}(x,x',\omega)&=\frac{{\rm e}^{i(k_{e}x-k_{h}x')}\,m_{5}-{\rm e}^{-i(k_{e}x-k_{h}x')}\,m_{6}}{2}\,,\\
\end{split}
\end{equation}
and $f^r_{1,N} = f^r_{2,N} = 0$, where $m_{5,6}$ are given by Eqs.\,(\ref{mSNSApp}).

We can explicitly check the antisymmetry condition of these expressions.
For example, for the singlet component $f_{0,{\rm N}}^{r}(x,x',\omega)$ we have
\begin{equation}
\begin{split}
f_{0,{\rm N}}^{a}(x',x,-\omega)&=\frac{{\rm e}^{-i(k_{e}x-k_{h}x')}\,m_{7}^{*}(-\omega)+{\rm e}^{i(k_{e}x-k_{h}x')}\,m_{8}^{*}(-\omega)}{2}\,,\\
&=\frac{{\rm e}^{i(k_{e}x-k_{h}x')}\,m_{5}+{\rm e}^{-i(k_{e}x-k_{h}x')}\,m_{6}}{2}\,,\\
&=f_{0,{\rm N}}^{r}(x,x',\omega)\,,
\end{split}
\end{equation}
and therefore we conclude that $f_{0,{\rm N}}^{r}(x,x',\omega)$ is antisymmetric according to the relations in Eqs.\,(\ref{antif}).
Likewise, we have verified the antisymmetry condition $f^{r}_{3,N}(x,x',\omega)=-f_{3,N}^{a}(x',x,-\omega)$. 
By using Eq.\,(\ref{Odd_Even_text}) we write down the even- and odd-frequency components
\begin{equation}
\label{SNS_N}
\begin{split}
f_{0,{\rm N}}^{r,{\rm E}}(x,x',\omega)&=W_{+}(\omega){\rm cos}[k_{\mu}(x-x')]\,,\\
f_{0,{\rm N}}^{r,{\rm O}}(x,x',\omega)&=W_{-}(\omega)i{\rm sin}[k_{\mu}(x-x')]\,,\\
f_{3,{\rm N}}^{r,{\rm E}}(x,x',\omega)&=W_{+}(\omega)i{\rm sin}[k_{\mu}(x-x')]\,,\\
f_{3,{\rm N}}^{r,{\rm O}}(x,x',\omega)&=W_{-}(\omega){\rm cos}[k_{\mu}(x-x')]
\end{split}
\end{equation}
where $W_{\pm}(\omega)=[m_{5}{\rm e}^{i(x+x')/\xi_{\omega}}\pm m_{6}{\rm e}^{-i(x+x')/\xi_{\omega}} ]/2$. 
These equations correspond to Eqs.\,(\ref{SNS_N_main}) presented in the main text. We have checked that Eqs.\,(\ref{SNS_N}) fulfill the antisymmetry conditions given by Eqs.\,(\ref{antif}). 
\begin{widetext}

\subsubsection{Superconducting regions}
In the left superconducting region we obtain
\begin{equation}
\label{SNS_G_L}
\begin{split}
G^{r}_{ee,\uparrow\uparrow}(x,x',\omega)&=Z
\Big[
a_{1}(\omega,L_{\rm N},\phi){\rm e}^{i(k_{h}^{S}x-k_{e}^{S}x')}+P(x,x',\omega)\Big]\,,\\
G^{r}_{ee,\downarrow\downarrow}(x,x',\omega)&=Z
\Big[a_{2}(\omega,L_{\rm N},\phi){\rm e}^{i(k_{h}^{S}x'-k_{e}^{S}x)}+P(x',x,\omega)\Big]\,,\\
G^{r}_{hh,\downarrow\downarrow}(x,x',\omega)&=Z
\Big[
a_{1}(\omega,L_{\rm N},\phi){\rm e}^{i(k_{h}^{S}x-k_{e}^{S}x')}+Q(x,x',\omega)\Big]\\
G^{r}_{hh,\uparrow\uparrow}(x,x',\omega)&=Z
\Big[a_{2}(\omega,L_{\rm N},\phi){\rm e}^{i(k_{h}^{S}x'-k_{e}^{S}x)}+Q(x',x,\omega)\Big]\,,\\
G^{r}_{eh,\uparrow\downarrow}(x,x',\omega)&=Z
\Big[
a_{1}(\omega,L_{\rm N},\phi)\frac{v}{u}{\rm e}^{i(k_{h}^{S}x-k_{e}^{S}x')}+\bar{P}(x,x',\omega)\Big]\\
G^{r}_{eh,\downarrow\uparrow}(x,x',\omega)&=Z
\Big[a_{2}(\omega,L_{\rm N},\phi)\frac{u}{v}{\rm e}^{i(k_{h}^{S}x'-k_{e}^{S}x)}+\bar{P}(x',x,\omega)\Big]\,,\\
G^{r}_{he,\downarrow\uparrow}(x,x',\omega)&=Z
\Big[
a_{1}(\omega,L_{\rm N},\phi)\frac{u}{v}{\rm e}^{i(k_{h}^{S}x-k_{e}^{S}x')}+\bar{Q}(x,x',\omega)\Big]\,,\\
G^{r}_{he,\uparrow\downarrow}(x,x',\omega)&=Z
\Big[a_{2}(\omega,L_{\rm N},\phi)\frac{v}{u}{\rm e}^{i(k_{h}^{S}x'-k_{e}^{S}x)}+\bar{Q}(x',x,\omega)\Big]\,,
\end{split}
\end{equation}
where  
\begin{displaymath}
\begin{split}
P(x,x',\omega)&=\theta(x-x')\frac{u}{v}{\rm e}^{ik_{e}^{S}(x-x')}+\theta(x'-x)\frac{v}{u}{\rm e}^{ik_{h}^{S}(x-x')}\,,\\
 Q(x,x',\omega)&=\theta(x-x')\frac{v}{u}{\rm e}^{ik_{e}^{S}(x-x')}
+\theta(x'-x)\frac{u}{v}{\rm e}^{ik_{h}^{S}(x-x')}\,, \\
\bar{P}(x,x',\omega)&=\Big[\theta(x-x'){\rm e}^{ik_{e}^{S}(x-x')}
+\theta(x'-x){\rm e}^{ik_{h}^{S}(x-x')}\Big]\,,\\
 \bar{Q}(x,x',\omega)&=\Big[\theta(x-x'){\rm e}^{ik_{e}^{S}(x-x')}
+\theta(x'-x){\rm e}^{ik_{h}^{S}(x-x')}\Big]\,,\\
\end{split}
\end{displaymath}
with $Z=\frac{1}{iv_{f}}\frac{1}{(u/v)-(v/u)}$  and $\bar{Z}=\frac{1}{-iv_{f}}\frac{1}{(u/v)^{*}-(v/u)^{*}}$.
In this case the Andreev reflection amplitudes obey  $a_{2}(\omega,L_{\rm N},\phi)=a_1(\omega,L_{\rm N},-\phi)$ 
with $a_{1}(\omega,L_{\rm N},\phi)$ given by Eqs.\,(\ref{ARCOEFFI}).
This results in the pairing amplitudes
\begin{equation}
\begin{split}
f^{r}_{0,{\rm S_L}}(x,x',\omega)&=
\frac{Z}{2}{\rm e}^{-ik(\omega)(x+x')}\Big[
a_{1}(\omega,L_{\rm N},\phi)\frac{v}{u}{\rm e}^{ik_{\mu}(x-x')}+a_{2}(\omega,L_{\rm N},\phi)\frac{u}{v}{\rm e}^{-ik_{\mu}(x-x')}\Big] +Z{\rm e}^{ik(\omega)|x-x'|}C_{xx'}\,,\\
f^{r}_{3,{\rm S_L}}(x,x',\omega)&=
\frac{Z}{2}{\rm e}^{-ik(\omega)(x+x')}\Big[
a_{1}(\omega,L_{\rm N},\phi)\frac{v}{u}{\rm e}^{ik_{\mu}(x-x')}-a_{2}(\omega,L_{\rm N},\phi)\frac{u}{v}{\rm e}^{-ik_{\mu}(x-x')}\Big] +Z{\rm e}^{ik(\omega)|x-x'|}iS_{xx'}\,,\\
\end{split}
\end{equation}
and $f^r_{1,{\rm S_L}} = f^r_{2,{\rm S_L}} = 0$.
In a similar way we proceed for the right superconducting region to calculate the pairing amplitudes.
We can then extract the odd- and even-frequency components. For energies within the superconducting gap we obtain
\begin{equation}
\label{SNS_L}
\begin{split}
f_{0,\pm}^{r,{\rm E}}(x,x',\omega)&=Z{\rm e}^{i\phi(1\mp1)/2}\Big\{{\rm e}^{\pm\kappa(\omega)(x+x')}
\Big[a_{1(4)}(\omega,L_{\rm N},\phi)a_{1}(\omega)+a_{2(3)}(\omega,L_{\rm N},\phi)a_{1}^{*}(\omega)\Big]+2{\rm e}^{-\kappa(\omega)|x-x'|}\Big\}C_{xx'}\,,\\
f_{0,\pm}^{r,{\rm O}}(x,x',\omega)&=Z{\rm e}^{i\phi(1\mp1)/2}\Big\{{\rm e}^{\pm\kappa(\omega)(x+x')}
\Big[a_{1(4)}(\omega,L_{\rm N},\phi)a_{1}(\omega)-a_{2(3)}(\omega,L_{\rm N},\phi)a_{1}^{*}(\omega)\Big]\Big\}iS_{xx'}\,,\\
f_{3,\pm}^{r,{\rm E}}(x,x',\omega)&=Z{\rm e}^{i\phi(1\mp1)/2}\Big\{{\rm e}^{\pm\kappa(\omega)(x+x')}
\Big[a_{1(4)}(\omega,L_{\rm N},\phi)a_{1}(\omega)+a_{2(3)}(\omega,L_{\rm N},\phi)a_{1}^{*}(\omega)\Big]+2{\rm e}^{-\kappa(\omega)|x-x'|}\Big\}iS_{xx'}\,,\\
f_{3,\pm}^{r,{\rm O}}(x,x',\omega)&=Z{\rm e}^{i\phi(1\mp1)/2}\Big\{{\rm e}^{\pm\kappa(\omega)(x+x')}
\Big[a_{1(4)}(\omega,L_{\rm N},\phi)a_{1}(\omega)-a_{2(3)}(\omega,L_{\rm N},\phi)a_{1}^{*}(\omega)\Big]\Big\}C_{xx'}\,,
\end{split}
\end{equation}
where $\pm$ subscripts correspond to results for the left and right superconducting regions, respectively.
Here we have used the notation $a_{1}(\omega)=\frac{v}{u}={\rm e}^{-i\eta(\omega)}$ and
$a_{1}(\omega,L_{\rm N},\varphi)$ given by Eq.\,(\ref{ARCOEFFI}).
Moreover, we find that $a_{3,4}(\omega,L_{\rm N},\varphi)=a_{2,1}(\omega,L_{\rm N},\varphi){\rm e}^{2\kappa(\omega)}$.
As with NS junctions, we associate bulk behavior to elements in the pairing amplitudes that do not depend on the Andreev reflections and have an exponential form ${\rm e}^{-\kappa(\omega)|x-x'|}$. We therefore conclude that the second term in the curly brackets in the ESE and ETO pairing amplitudes emerge in the bulk of S. 
However, at the interface, we observe all different allowed pairing amplitudes. 
Notice that all the pairing amplitudes in the right superconductor acquire a phase factor ${\rm e}^{i\phi}$.
\end{widetext}
\end{document}